\documentclass[longauth]{aa}
\usepackage[utf8]{inputenc}

\usepackage{graphicx}
\usepackage{rotating,booktabs}
\usepackage[verbose]{placeins}
\usepackage{pdflscape}
\usepackage{adjustbox}
\usepackage{tikz}
\usepackage{enumitem}
\usepackage[english]{babel}
\usepackage{url}
\usepackage{hyperref}
\hypersetup{colorlinks   = true, citecolor = blue,
            urlcolor=blue,
            linkcolor=blue}

\usepackage{txfonts}
\usepackage{natbib}
\usepackage{longtable}
\usepackage{amsmath}
\usepackage{mhchem}
\newcommand{\ha}{\ce{H$\alpha$}}
\newcommand{\hb}{\ce{H$\beta$}}
\newcommand{\sii}{[\ion{S}{II}]}
\newcommand{\nii}{[\ion{N}{II}]}
\newcommand{\naid}{\ion{Na}{ID}}

\newcommand{\oi}{[\ion{O}{I}]}
\newcommand{\oiii}{[\ion{O}{III}]}
\newcommand{\oiiism}{[\ion{O}{III}]$\lambda 4960$\AA}
\newcommand{\oiiihi}{[\ion{O}{III}]$\lambda 5008$\AA}
\newcommand{\niifr}{[\ion{N}{II}]$\lambda 6548$\AA}
\newcommand{\niil}{[\ion{N}{II}]$\lambda 6583$\AA}

\newcommand{\cott}{\ion{CO}{(3-2)}}
\newcommand{\coo}{\ion{CO}{(1-0)}}

\newcommand{\feii}{\ion{Fe}{II}}
\newcommand{\civ}{\ion{C}{IV}}
\newcommand{\vs}{$v_{\rm{shift}}$}
\newcommand{\sv}{$\sigma_{\rm{vel}}$}
\newcommand{\mdotout}{$\dot{M}_{\rm{out}}$}

\newcommand{\edotkin}{$\dot{E}_{\rm kin}$}
\newcommand{\pdotout}{$\dot{P}_{\rm out}$}
\defcitealias{Bischetti19}{B19}
\defcitealias{Fiore17}{F17}

\begin{document} 

\title{MUSE view of PDS 456: kpc-scale wind, extended ionized gas and close environment}

\authorrunning{A. Travascio et al.}
\titlerunning{MUSE view of PDS 456: kpc-scale wind, extended ionized gas and close environment}

\author{A. Travascio
          \inst{1,2},
          E. Piconcelli
          \inst{3},
          M. Bischetti
          \inst{2,4},
          G. Cresci
          \inst{5},
          C. Feruglio
          \inst{2},
          M. Perna
          \inst{6},
          G. Vietri
          \inst{7},
          S. Carniani
          \inst{8},
          S. Cantalupo
          \inst{1},
          C. Cicone
          \inst{9},
          M. Ginolfi
          \inst{10},
          G. Venturi
          \inst{8,5},
          K. Zubovas
          \inst{11,12},
          A. Bongiorno
          \inst{3},
          M. Brusa
          \inst{13,14},
          A. Luminari
          \inst{15},
          V. Mainieri
          \inst{16},
          A. Marconi
          \inst{10},
          N. Menci
          \inst{3},
          E. Nardini
          \inst{5},
          A. Pensabene
          \inst{1},
          C. Ramos Almeida
          \inst{17,18},
          F. Tombesi
          \inst{3,19,20,21,22},
          C. Vignali
          \inst{13,14},
          L. Zappacosta
          \inst{3}
          and
          F. Fiore
          \inst{2}
          }

\institute{Dipartimento di Fisica “G. Occhialini”, Università degli Studi di Milano-Bicocca, Piazza della Scienza 3, I-20126, Milano, Italy \\ e-mail: \texttt{andrea.travascio@unimib.it} \and 
   INAF – Osservatorio Astronomico di Trieste, Via G. B. Tiepolo 11, 34143 Trieste, Italy \and
   INAF – Osservatorio Astronomico di Roma, Via di Frascati 33, 00040 Monteporzio Catone, Rome, Italy \and
   Dipartimento di Fisica, Universitá di Trieste, Sezione di Astronomia, Via G.B. Tiepolo 11, I-34131 Trieste, Italy \and
   INAF – Osservatorio Astrofisico di Arcetri, Largo E. Fermi 5, 50127 Firenze, Italy \and
   Centro de Astrobiología (CAB), CSIC–INTA, Cra. de Ajalvir Km. 4, 28850 Torrejón de Ardoz, Madrid, Spain \and 
   INAF - Istituto di Astrofisica Spaziale e Fisica cosmica Milano, Via Alfonso Corti 12, 20133, Milano, Italy \and
   Scuola Normale Superiore, Piazza dei Cavalieri 7, 56126 Pisa, Italy \and
   Institute of Theoretical Astrophysics, University of Oslo, P.O. Box 1029, Blindern 0315, Oslo, Norway \and
   Dipartimento di Astronomia e Scienza dello Spazio, Universià degli Studi di Firenze, Largo E. Fermi 2, 50125 Firenze, Italy \and
   Center for Physical Sciences and Technology, Saul\.{e}tekio al. 3, Vilnius LT-10257, Lithuania \and 
   Astronomical Observatory, Vilnius University, Saul\.{e}tekio al. 3, Vilnius LT-10257, Lithuania \and 
   Dipartimento di Fisica e Astronomia "Augusto Righi", Università degli Studi di Bologna, Via P. Gobetti, 93/2, 40129 Bologna, Italy \and 
   INAF-Osservatorio di Astrofisica e Scienza dello Spazio di Bologna, Via Piero Gobetti, 93/3, 40129 Bologna, Italy \and 
   INAF – Istituto di Astrofisica e Planetologia Spaziali,Via del Fosso del Caveliere 100,00133 Roma,Italy \and 
   European Southern Observatory, Karl-Schwarzschild-Strasse 2, Garching bei München, Germany \and 
   Instituto de Astrofísica de Canarias, Calle Vía Láctea, s/n, 38205 La Laguna, Tenerife, Spain \and 
   Departamento de Astrofísica, Universidad de La Laguna, 38206 La Laguna, Tenerife, Spain \and 
   Dept. of Physics, University of Rome “Tor Vergata”, Via della Ricerca Scientifica 1, 00133 Rome, Italy \and 
    Dept. of Astronomy, University of Maryland, College Park, MD 20742, USA \and 
    NASA – Goddard Space Flight Center, Code 662, Greenbelt, MD 20771, USA \and 
    INFN – Roma Tor Vergata, Via Della Ricerca Scientifica 1, 00133 Rome, Italy
   }

\abstract{PDS 456 is the most luminous ($L_{\rm bol}\sim 10 ^{47}~\rm{erg ~s^{-1}}$) radio-quiet quasar at $z<0.3$ and can be regarded as a local counterpart of the powerful quasars shining at Cosmic Noon. It hosts a strong nuclear X-ray ultra-fast ($\sim 0.3c$) outflow, and a massive and clumpy \cott\ molecular outflow extending up to $\sim$5~kpc from the nucleus.
We analyzed the first MUSE Wide Field Mode (WFM) and Adaptive-Optics Narrow Field Mode (AO-NFM) optical integral field spectroscopic observations of PDS456. The AO-NFM observations provide an unprecedented spatial resolution, reaching up to $\sim$280 pc. 
Our findings reveal a complex circumgalactic medium around PDS 456, extending up to a maximum projected size of $\approx$46 kpc. This includes a reservoir of gas with a mass of $\sim 10^7-10^8 \rm ~M_{\odot}$, along with eight companion galaxies, and a multi-phase outflow. WFM and NFM MUSE data reveal an outflow on a large scale ($\approx$12 kpc from the quasar) in \oiii , and on smaller scales (within 3 kpc) with higher resolution (about 280 pc) in \ha , respectively. 
The \oiii\ outflow mass rate is $\rm 2.3 \pm 0.2 ~M_{\odot}~yr^{-1}$ which is significantly lower than those typically found in other luminous quasars. Remarkably, the \ha\ outflow shows a similar scale, morphology, and kinematics to the \cott\ molecular outflow, with the latter dominating in terms of kinetic energy and mass outflow rate by two and one orders of magnitude, respectively.
Our results therefore indicate that mergers, powerful AGN activity, and feedback through AGN-driven winds will collectively contribute to shaping the host galaxy evolution of PDS 456, and likely, that of similar objects at the brightest end of the AGN luminosity function across all redshifts.
Moreover, the finding that the momentum boost of the total outflow deviates from the expected energy-conserving expansion for large-scale outflows highlights the need of novel AGN-driven outflow models to comprehensively interpret these phenomena.}

   \keywords{Galaxies: nuclei --
             Galaxies: ISM --
             Galaxies: interactions --
             quasars: individual: PDS 456 --
             quasars: emission lines 
             }

\maketitle
%

\section{Introduction} \label{sec:intro}

Most of the massive galaxies host a supermassive black hole (SMBH) in their center, that accrete material during a phase called Active Galactic Nucleus (AGN), releasing a large amount of energy affecting the host-galaxy itself. We call this phenomenon ``AGN feedback''. SMBHs reach their maximum accreting and feedback efficiency at $z \sim 2-3$, which we refer to as ``Cosmic Noon'', i.e. the peak epoch of galaxy assembly and SMBH accretion \citep[see][and references therein]{Schreiber20}. 
Understanding AGN feedback is essential for explaining key observational findings like SMBH-galactic correlations \citep{Magorrian98}, star formation inefficiency at high-$\rm M_{\star}$ \citep{Binney95}, and intergalactic medium enrichment \citep{Tytler95}.

During the AGN phase, the released energy triggers semi-relativistic nuclear winds via radiative pressure-driven \citep{Proga98,Proga00} or magnetic pressure-driven \citep{Fukumura10,Luminari21} mechanisms, leading to X-ray \citep[Ultra Fast Outflow; UFO][]{Reeves03} and UV broad absorption line \citep{Jannuzi96,Vietri22}. These winds shock the Inter-Stellar Medium (ISM) gas, generating galactic-scale, multi-phase outflows \citep{King05,Lapi05,Menci08,Faucher12}. 
 
If the energy of the shocked wind is radiated away and the momentum is conserved \citep[momentum-conserving scenario;][]{Fabian99,King03}, outflows are expected to extend up to a few hundred parsecs \citep{Feruglio17,Zanchettin21}. An energy-conserving scenario is proposed to explain the expansion of kiloparsec-scale outflows \citep[][hereafter \citetalias{Fiore17}]{Silk98,King05,Zubovas12,Nayakshin14,Fiore17}. In this context, the nuclear wind undergoes adiabatic expansion, facilitated by a cooling time for the post-shock outflow that surpasses its dynamical time, thus ensuring energy conservation. This scenario remains valid beyond a critical distance of more than $100~\rm pc$ from the quasar resulting in outflows exhibiting a momentum boost of $\sim$10-20 \citep{Zubovas14,King15}. However, in a two-temperature plasma scenario, where ions decouple thermally from electrons due to the shock, the cooling radius is much smaller than $100~\rm pc$ \citep{Faucher12}. Moreover, \cite{Zubovas14b} demonstrate that the efficiency in depositing energy into the ambient gas also depends on the clumpiness of the gas.

More recent observational investigations \citep{Bischetti19,Smith19,Reeves19,Sirressi19,Marasco20,Tozzi21,Speranza22,Bonanomi23} have revealed a more complex scenario, supporting a momentum-conserving scenario for both molecular and ionized galactic-scale outflows, at least in some cases. Still other AGN driven large-scale outflows show momentum loading factors that fall below even the momentum-conserving prediction, while others have extremely high factors, above the energy-driven expectations \citep{Marasco20}. Several hypotheses have been put forward to explain these observations. One scenario predicts that galactic-scale outflows could be primarily driven by the radiative pressure from AGN photons on dust \citep{Fabian12,Ishibashi18a,Costa18}. 
This mechanism can account for galactic-scale outflows maintaining a low momentum boost. It predicts a super-linear correlation (i.e. correlation with a slope greater than one) between the kinetic power of large-scale outflows and AGN luminosity; such a correlation is supported by outflow observations (e.g.,  \citetalias{Fiore17}, \citealt{Fluetsch21}, \citealt{Lutz20}). On the other hand, this model does not explain the spread of outflow properties in different AGN with the same luminosity.
Another possible explanation relies on considering an intermittent AGN luminosity history \citep[see][]{Nardini18,Zubovas20,Zubovas22}. These papers rely entirely on the energy-driven outflow paradigm but show that outflow properties correlate much more strongly with the long-term (several Myr) average AGN luminosity rather than the instantaneous one. The observed momentum and energy loading factors are then determined to a large extent by the ratio of the current AGN luminosity to the long-term average.\\


In this paper, we use MUSE Adaptive Optics (AO) -assisted Narrow Field Mode (NFM), and Wide Field Mode (WFM) observations to explore the properties of the environment and the ionized outflow in PDS 456, that is the most luminous (bolometric luminosity; $L_{\rm bol} \sim 10^{47} ~\rm{erg~s^{-1}}$) radio-quiet quasar at $z<0.3$ \citep[$z_{\rm CO} = 0.185$;][hereafter \citetalias{Bischetti19}]{Bischetti19}, discovered by \cite{Torres97} in the Pico dos Dias survey. 
The analysis of the Spectral Energy Distribution (SED) reveals both the quasar-like and the ULIRG-like nature of PDS 456, suggesting that it is undergoing a transition from a Luminous IR Galaxy (LIRG) to a quasar \citep{Yun04}. 
In \citetalias{Bischetti19}, the star formation rate (SFR) of the host galaxy is estimated to be $\approx 30-80~\rm{M_{\odot}~yr^{-1}}$. 
In addition, \cite{Yun04} conducted an analysis using multiple datasets, including a 0.6$''$ resolution K-band image obtained from the Keck Telescope, a VLA continuum image at 1.2 GHz, and the CO(1-0) emission with the Owens Valley Radio Observatory (OVRO) millimeter array. Their study revealed the presence of three compact continuum sources located $\approx 10~ \rm kpc$ southwest of the quasar. Furthermore, in their ALMA observations, over a region of $\rm \sim 50 \times 50~kpc^2$, \citetalias{Bischetti19} identified \coo -emitting companions near PDS 456. 

\cite{Simpson99} were the first to analyze the optical spectrum, revealing the presence of broad line region (BLR) Balmer lines (FWHM $> 4000 \rm ~km~ s^{-1}$), a weak \oiii\ emission line, and prominent \feii\ transitions.
PDS 456 can be seen as the local counterpart of the luminous quasars shining at $z \gtrsim 2$, where we expect a high efficiency in driving radiative winds \citep[e.g.][]{Brusa15,Carniani15,Bischetti17,Forster18,Kakkad20}. Indeed, PDS 456 exhibits a relativistic ($\sim 0.3 c$) and powerful ultra fast X-ray winds (kinetic power; $\dot{E}_{\rm{kin}} \simeq 0.2 L_{\rm{bol}}$), detected through Fe K-shell absorption features \citep{Reeves09,Nardini15,Luminari18}. 
\cite{Hamann18} reported the possible presence of a fast ($\sim 0.3 c$) UV broad absorption line wind in \civ, exhibiting similar velocities to the nuclear X-ray winds. Additionally, \cite{OBrien05} measured a $\sim 5000~\rm  km~s^{-1}$ blueshift of the \civ\ emission line with respect to the systemic redshift, likely associated with an outflow in the quasar broad line region.
The presence of a molecular outflow was detected in ALMA data by \citetalias{Bischetti19}. They observed a blueshifted ($< -250 \rm ~ km ~s^{-1}$) \cott\ outflow extending eastward from the quasar, with a maximum projected distance of $\sim$5 kpc. Additionally, a more compact ($<1 ~\rm kpc$) west-oriented outflow was observed, exhibiting low positive velocities ($\sim 80 - 150~\rm km ~s^{-1}$).
The molecular outflow has a low momentum boost (i.e. $\dot{P}_{\rm{out}}/\dot{P}_{\rm{rad}} \approx 0.36$, where $\dot{P}_{\rm{out}} = v_{\rm{max}} \times \dot{M}_{\rm{out}}$, with \mdotout\ the mass rate of the outflow and $v_{\rm{max}}$ the maximum velocity of the outflow, and $\dot{P}_{\rm{rad}} = L_{\rm bol}/c$) which is not consistent with an energy-conserving scenario. 
Indeed, in $log(L_{\rm bol}/\rm erg~s^{-1}) \gtrsim 47$ quasars, the energetic contribution of the ionized outflows is found to be comparable to that of the molecular phase, thus $\dot{M}_{\rm out,ion} \sim \dot{M}_{\rm out,mol}$ for $L_{\rm bol} = 10^{47}~\rm{erg~s^{-1}}$ \citepalias{Fiore17}.
Nevertheless, although the mass outflow rate measured for the molecular outflow ($\dot{M}_{\rm out} \sim 290 \rm ~M_{\odot}~yr^{-1}$) is well below the prediction from the empirical relation by \citetalias{Fiore17}, it is sufficient to deplete the molecular gas reservoir in $\rm \sim 10~Myr$, leading to a rapid suppression of the star formation activity.
 
Finally, although PDS 456 is classified as a radio-quiet object, \cite{Yang21} recently reported the detection of a faint ($L_{1.66~\rm{GHz}} < 10^{40.2} ~\rm{erg~s^{-1}}$), complex radio structure. This consists of collimated jets, and an extended (up to 360 pc) non-thermal radio emission, which is explained as shock emission due to the interaction between nuclear winds and high-density regions of the ISM. The energy associated with these radio structures ($L_R/L_{\rm{bol}} = 10^{-7}$) is significantly lower compared to that of the nuclear winds \citep{Nardini15,Luminari18}.

In conclusion, PDS 456 is an excellent laboratory to study the environment and the feeding and feedback processes involved in the typical scenario of efficient AGN phase observed at Cosmic Noon around hyper-luminous quasars.
In particular, the large FoV of the MUSE WFM Integral Field Spectroscopy (IFS) data is crucial to study the environment of this system traced by the ionized phase, i.e. the Narrow Line Region (NLR), host-galaxy, companions and diffuse emission.
On the other hand, the MUSE AO-NFM data of the PDS 456 center provides observations with unprecedented spatial-resolution ($280~\rm pc$) for this type of system, with the aim to study the mechanisms of expansion of any possible ionized-phase outflow associated with the molecular one detected with ALMA in \citetalias{Bischetti19}.\\

Sect.~\ref{sec:reductionmuse} describes the MUSE WFM and NFM observations, while the analysis of these data and the corresponding results are reported in Sects.~\ref{sec:WFMobs} and \ref{sec:NFMobs}, respectively. In these sections, we investigate the environment of PDS 456, as well as the morphology and kinematics of the extended emitting gas and the presence of companion galaxies. 

Sect.~\ref{sec:Energy} explores the physical and energetic properties of the outflow detected in both observation modes across different scales. 
Finally, in Sect.~\ref{sec:expansionOut}, we discuss possible scenarios for the expansion of these outflows based on their properties. A summary of the main results is provided in Sect.~\ref{sec:summconc}.\\

To be consistent with \citetalias{Bischetti19}, we consider the systemic redshift of PDS 456 of $z=0.185$, as well as a physical scale of 3.126 kpc/arcsec at the following cosmological parameters: $H_0 = 69.6~\rm km~s^{-1}~Mpc^{-1}$, $\Omega_m=0.286$, and $\Omega_{\Lambda}=0.714$.

\begin{figure*}[t]
   \begin{center}
   \includegraphics[height=0.4\textheight,angle=0]{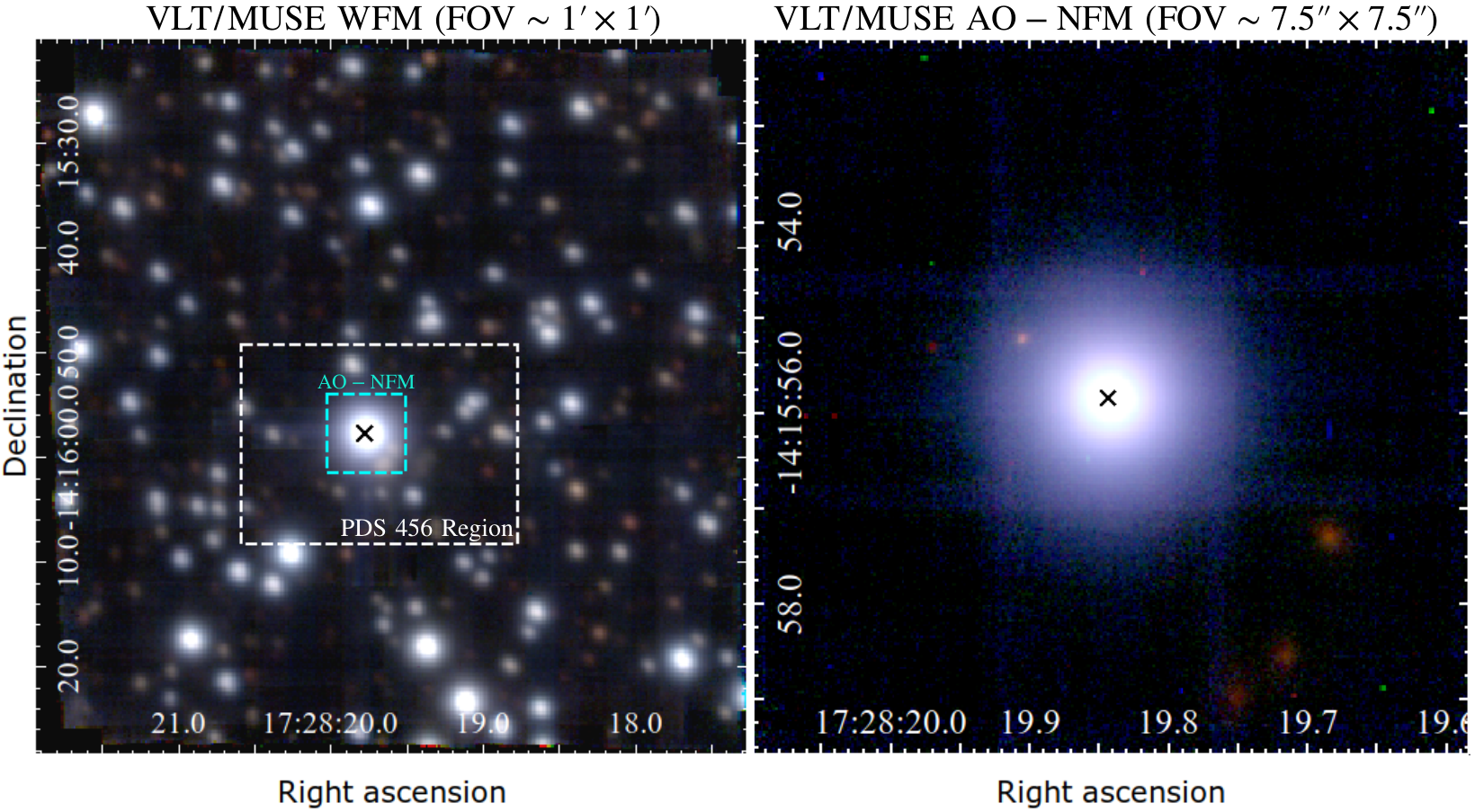}
   \caption{RGB wide filters images of the WFM (left) and NFM (right) MUSE datacube, obtained by collapsing the MUSE datacubed in the same spectral regions of the HST/ACS-HCR filters F475W, F625W and F775W for the WFM data, and F459M, F660N and F892N for the NFM data. For the NFM data, spectral regions contaminated by instrument features fallen in the filters are excluded. In the WFM image, we highlight the region studied in this work, including PDS 456 and companions, with a dashed white box measuring $\rm 80 \times 60~kpc^2$. The cyan box indicates the size ($\approx 23 \times 23~\rm kpc^2$) of the NFM FOV centered on PDS 456. The positions of the nucleus of PDS 456 are marked with black crosses.}
   \label{fig:WLImages}
   \end{center}
\end{figure*}


\section{Data Reduction}\label{sec:reductionmuse}

In this paper we analyze MUSE IFS observations in WFM and in AO-NFM taken on April to June 2019 (PI: E. Piconcelli). The MUSE WFM observations consisted of four Observing Blocks (OBs) of 600 sec each, for a total exposure time of 1 h. Each exposure was rotated by 90 deg with the addition of a small dithering. 
Instead, the MUSE AO-NFM observations consist of three OBs for a total observing time of $\sim$3 h, and each OB consists of 8 exposures including sky pointings. 
The observations were acquired with seeing and airmass $\sim$0.4''-1'' and $\sim$1.1, respectively.

Data were reduced using the ESO MUSE standard pipeline with EsoRex v. 3.12.3 \citep{Weilbacher14}. The sky frame for MUSE WFM raw data was generated directly from the ``OBJECT'' exposures using the standard ESO pipeline. For AO-NFM raw data, the sky frame was created from the pixel tables of exposures taken in empty sky regions, i.e. the ``SKY'' exposures.

The final WFM (NFM) datacube is characterized by a FoV of $ \approx 1' \times 1'$ ($\approx$7$\times$7 $\rm arcsec^2$), and a pixel size of 0.2 (0.025) arcsec, that is equivalent to $\sim$625 pc ($\sim$78 pc) at $z=0.185$. The FWHM of the final datacubes' PSF are $\sim 1''$ (i.e. $\sim 3~\rm kpc$) and $\sim 0.09''$ (i.e. $\sim 280~\rm pc$) for WFM and NFM observations, respectively. These values are estimated based on the individual point-sources within the fields. The spectral range covered is from 4750 \AA\ to 9350 \AA, with a spectral bin of 1.25 \AA. 
However, the presence of the \naid\ notch filter, which is required for the laser guide star in the AO mode, significantly contaminates the spectral region ranging from approximately 5500 to 6000~\AA\ in the MUSE NFM data. Consequently, we did not utilize this contaminated portion of the data. We verified the absolute wavelength calibration in our final datacube by checking the positions of the most intense sky lines. Finally, we checked the calibration of the flux in NFM-MUSE data comparing the total PSF of PDS 456 with that in WFM-MUSE data as reference. 

Fig~\ref{fig:WLImages} shows the RGB images of PDS 456.
In the left panel, we present an image of the entire FoV from the WFM-MUSE data. To create this image, we overlay mimic-wide filters images that are generated by collapsing the spectrum of the datacube into three spectral regions simulating the filters in HST/ACS-HCR: F475W (B), F625W (G) and F775W (R). This allows us to obtain a rough approximation of the color of each source within the FoV. 
The right panel displays the NFM MUSE observation, presented as an RGB image overlaid with mimic-wide filters images collapsed in the spectral regions of the HST/ACS-HCR filters F459M (B), F660N (G) and F892N (R). 

We use the positions of a handful of point sources identified by GAIA within the WFM MUSE observation field to refine the astrometry of our data. The astrometry of NFM MUSE data is automatically adjusted through alignment with the peak of the continuum emission from PDS 456.

\begin{figure}[t]
   \begin{center}
 \includegraphics[height=0.268\textheight,angle=0]{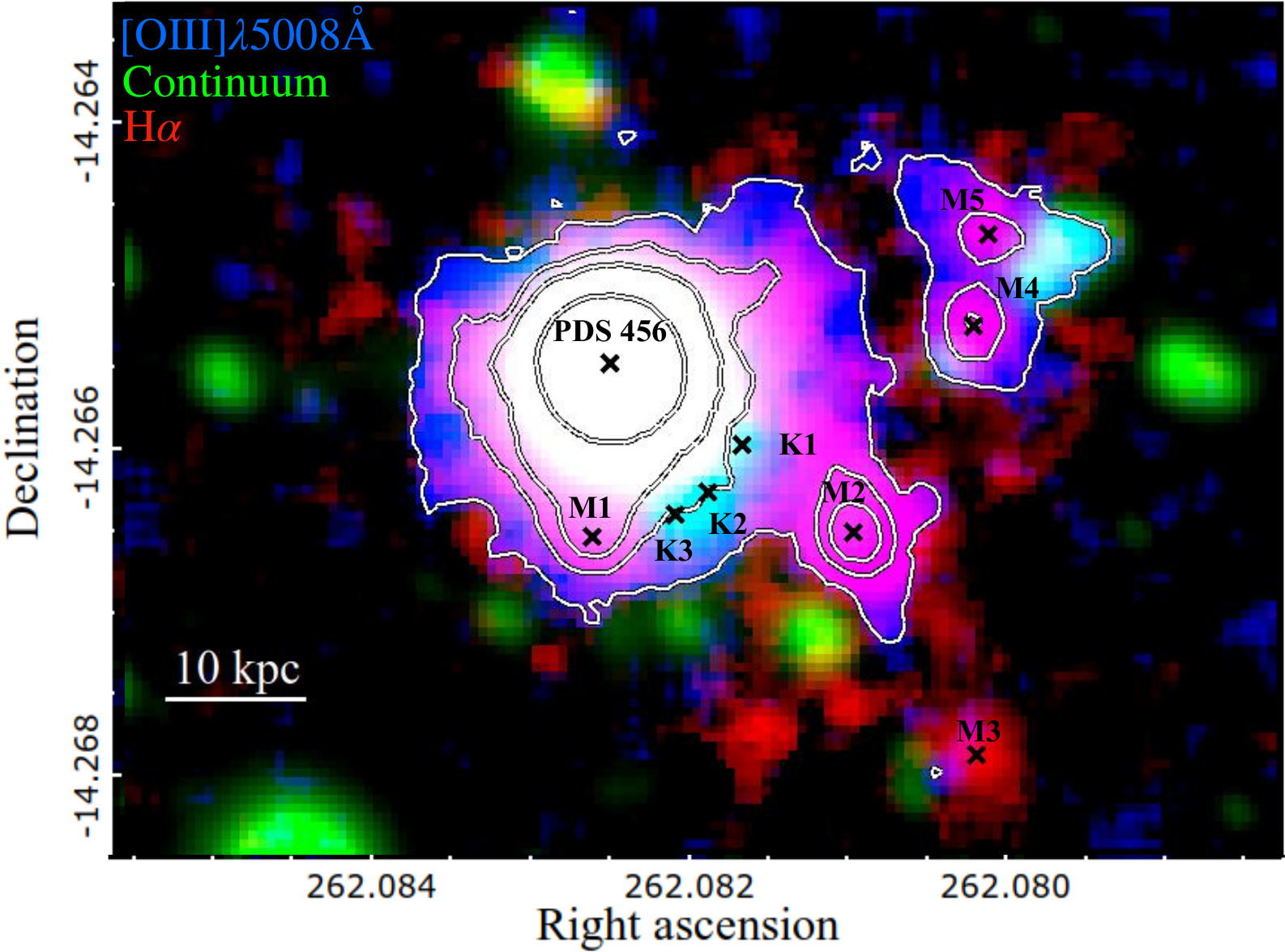}
   \caption{The extended emission surrounding PDS 456, within the region of the WFM-MUSE FoV indicated by a dashed white box in the left panel of Fig.~\ref{fig:WLImages}. This RGB composite image is generated from the following images: optimally-extracted narrow band images of \ha\ (R) and \oiiihi\ (B) obtained using the \texttt{CubEx} tool, and the White-Light Image (WLI; G). The black cross symbols mark the line-emitting companions detected with the Keck telescope \citep{Yun04}, ALMA \citepalias{Bischetti19}, and in this work with MUSE (see Table~\ref{tab:comp} for details). MUSE continuum sources are visible in green. The white contours show the surface brightness levels of the \oiii\ nebula at 10, 50, 100 and 500 $10^{-18}~ \rm{erg~s^{-1}~cm^{-2}~arcsec^{-2}}$.}
   \label{fig:WFMmaps}
   \end{center}
\end{figure}


\section{WFM observations}\label{sec:WFMobs}

\subsection{Subtraction of the Nuclear and Continuum Emission}\label{sec:WFMPSFSub}

We analyze the MUSE WFM observations with the purpose to investigate the properties of the kpc-scale diffuse gas around PDS 456 by subtracting the BLR emission, the iron transitions and the continuum emission (hereafter we refer to the combination of these emission components as "the nuclear component") from the region associated with the quasar PSF. For the WFM observations this task is complicated by the presence of several continuum sources blended with the quasar. For this reason we do not employ for the WFM the standard PSF subtraction tools used in literature (e.g., \texttt{CubePSFSub} see later for the discussion) and we develop a custom procedure as described in detailed below.

The initial step consists of extracting the spectrum from a small circular region centered on PDS 456 in the assumption that this area is completely dominated by the nuclear component. In order to derive a spectral model for this emission, we adopt an empirical approach since a purely analytical model results in large residuals, particularly from iron templates. The empirically calibrated model is obtained by performing a spline interpolation of the spectrum using third-degree polynomial after carefully masking the spectral regions containing narrow emission lines.
In order to pinpoint these spectral regions, we rely on a iterative procedure using visual inspection.
The model spectral shape is fixed while we vary the normalization for each spaxel through spectral fitting based on $\chi^2$ minimization after masking the spectral regions associated with narrow emission lines. Second order effects due to, e.g. wavelength PSF variations are accounted for with an additional spectral component modeled with a 7-degree polynomial. We stress that this additional component has minimal effects on our final result. Finally, the obtained spectral models are subtracted spaxel by spaxel. 

To evaluate any potential overestimation or underestimation of the narrow emission lines resulting from the subtraction of the best-fit nuclear component of PDS 456, we examine the \niil /\niifr\ and \oiiihi /\oiiism\ line ratios. Our analysis indicates that, within the $1 \sigma$ uncertainties, these ratios match with theoretical predictions \citep[$\approx 1/3$;][]{Storey00}.
As a further check we verify that the spatial extension of the nuclear component emission matches with the PSF estimated from some point sources in the WFM MUSE field.
As a final step, we follow the same procedure described in Sect 3.2 from \cite{Borisova16} (see also \citealt{Cantalupo19} for further details) in order to eliminate both any possible residual emission resulting from the subtraction of the nuclear component and any continuum sources across the entire WFM field of view.

\subsection{Morphology and extension of the ionized gas emission} \label{sec:ExtractionEmission}

\begin{table*}
\caption{List of companions detected around PDS 456. Col. (1) source name; col. (2) and (3): coordinates in degrees; col. (4) and (5): projected distances in kpc from PDS 456 and redshift, respectively; col. (6): tracers, such as continuous lines or bands, were employed to identify the redshift of the sources.}
\label{tab:comp}
\bigskip
\begin{adjustbox}{width=2\columnwidth,center} 
\begin{tabular}{c|c|c|c|c|c}
\hline
Src & RA & DEC & D(kpc) & $z$ & tracers  \\
(1) & (2) & (3) & (4) & (5) & (6)  \\
\hline
K1 &  262.0818 & -14.26599 &  9 & 0.1847$\rm ^a$ & Continuum K-band, Emission line CO(3-2), Absorption lines \naid\   \\
K2 & 262.0820 & -14.26629 & 10 & 0.1845$\rm ^a$ & Continuum K-band, Emission line CO(3-2), Absorption lines \naid\  \\
K3 & 262.0821 & -14.26640 & 11 & 0.1837$\rm ^a$ & Continuum K-band, Emission line CO(3-2), Absorption lines \naid\  \\
M1 &   262.0826  & -14.2665 & 11 & 0.1847 & Rest-frame optical emission lines and continuum in MUSE   \\
M2 &  262.0809  & -14.2666 & 21 & 0.1829/0.1836$\rm ^b$ & Emission line CO(3-2), rest-frame optical emission lines and continuum in MUSE  \\
M3 &   262.0802  & -14.2679 & 38 & 0.1838 & Rest-frame optical emission lines and continuum in MUSE   \\  
M4 &  262.0802  & -14.2653 & 25 & 0.1843 & Rest-frame optical emission lines and continuum in MUSE  \\
M5 &   262.0800  & -14.2647 & 28 & 0.1840 & Rest-frame optical emission lines and continuum in MUSE   \\  
\hline 
\end{tabular}
\end{adjustbox}
\footnotesize{The sources have been detected with ALMA observations in \citetalias{Bischetti19}, with the Near-Infrared Camera on the W. M. Keck Telescope in \cite{Yun04}, and with the IFS MUSE-WFM data in this work. $\rm ^a$ The redshift is obtained by fitting the \naid\ absorption feature. $\rm ^b$ The first (second) value indicates the \coo\ (\ha) based redshift;}
\end{table*}

To obtain a 3D map of the ionized gas emission, we use the \texttt{CubExtractor} tool on the datacube after subtracting the nuclear component and the continuum. This tool generates a 3D-mask comprising a minimum number of connected voxels (i.e. spatial and spectral elements; \texttt{MinNVox}=10000) above a user-defined threshold, and whose line emission is above a minimum signal-to-noise ratio (\texttt{SN\_Threshold}=3). These values have been widely used in the literature \citep[e.g.,][]{Borisova16,Battaia19,Cantalupo19} to extract extended emission.
This method allows us to trace the circumgalactic medium (CGM) in PDS 456 through the extended emission in the \hb, \ha, \niifr, \niil\, \oiii\ and \sii\ optical transitions.
We produce the optimally-extracted Narrow Band (NB) images for each detected emission line. These are images obtained by collapsing the only voxels in which the signal-to-noise ratio (SNR) of the emission line is above \texttt{SN\_Threshold} we set.

Fig.~\ref{fig:WFMmaps} shows an RGB image produced by including optimally-extracted NB images of the extended \ha\ (red) and \oiii\ (blue) emission and the White Light Image (WLI) highlighting the continuum (green) of the original datacube. 
This image reveals a complex morphology of the CGM around PDS 456, extending up to a maximum projected size of $\approx$46 kpc.

\begin{figure}[t]
   \begin{center} \includegraphics[height=0.7\textheight,angle=0]{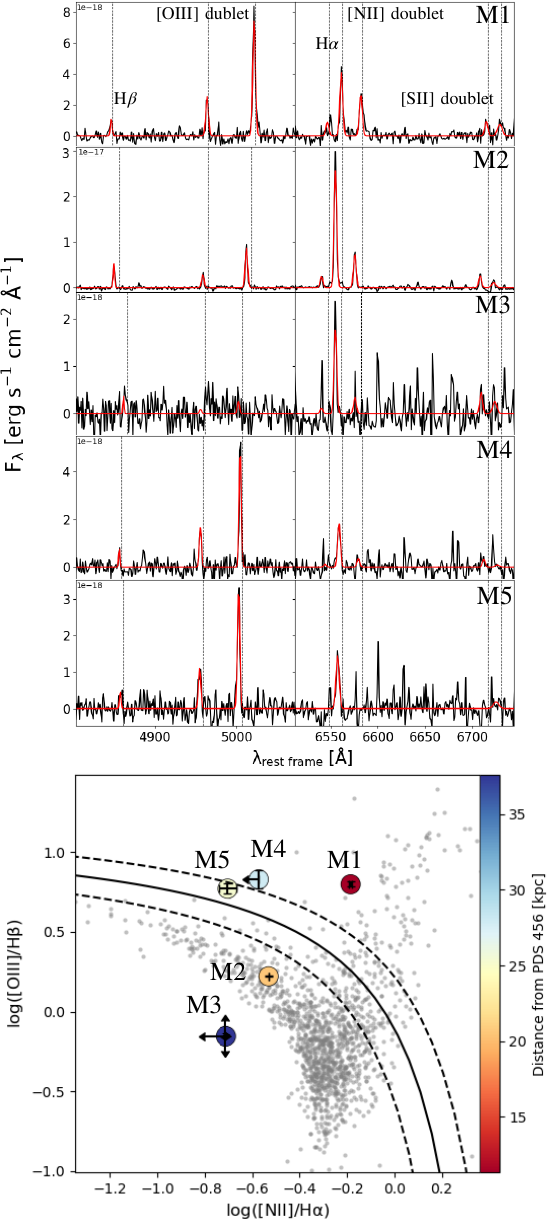}
   \caption{The top panel displays the spectra and their best-fit models, while the bottom panel presents the \nii -BPT diagram, for the galaxy companions of PDS 456. These spectra were extracted from circular regions with a radius of 2 pixels in the datacube, where the nuclear and continuum emission components of PDS 456 have been subtracted. Transparent gray dots show the values from the SDSS survey (credits: Jake Vanderplas $\&$ AstroML Developers), The black lines dividing different types of emissions in the BPT diagram were taken from \cite{Kewley01}.}
   \label{fig:BPTcompanions}
   \end{center}
\end{figure}

\subsection{Multiple Companions of PDS 456}\label{sec:companions}

The CGM emission shown in Fig.~\ref{fig:WFMmaps} also reveals the existence of components resembling bridges that connect various sources marked with black crosses. These sources share the same redshifts as PDS 456, confirmed through the detection of emission and absorption lines with ALMA (\citetalias{Bischetti19}), and with MUSE data in this study. A comprehensive list of companion sources, including their coordinates, distances from PDS 456, redshifts, and associated tracers, is provided in Table~\ref{tab:comp}.

Prominent continuum emission is observed from three sources near PDS 456, i.e. K1, K2, and K3, which were also detected in the K-band continuum using the Near-Infrared Camera on the Keck Telescope \citep{Yun04}. MUSE data reveal \naid\ absorption features, serving as proxies for neutral gas. The redshifts of these three sources are estimated by fitting the \naid\ absorption line using the model adopted in \cite{Sato09} and \cite{Perna20}. The best-fit models are presented in Fig.~\ref{figapp:abs} in Appendix~\ref{app:abs}. The redshifts derived in this way for K1 and K3 match with those obtained from \cott\ in \citetalias{Bischetti19}.
The absence of these \naid\ absorption features at the center of PDS456 is consistent with other luminous unobscured AGN \citep{Rupke05,VillarMartin14,Perna17,Perna19}. 
Regarding the other continuum sources highlighted in green in Fig.~\ref{fig:WFMmaps}, no distinctive features can be discerned to determine their redshifts. A more detailed analysis for this purpose is deferred to future investigations.
The MUSE spectra of five sources, i.e. M1--M5, show emission lines and and a less prominent continuum emission. Notably, the continuum emission for these sources is not visible in Fig.~\ref{fig:WFMmaps}, but it is observed in the extracted spectra.
Therefore, all eight sources listed in Table~\ref{tab:comp} exhibit both emission/absorption lines and continuum, identifying them as galaxy companions. Notably, the majority of companions are situated to the west of the central quasar at distances ranging from 9 to 38 kpc.

To determine the nature and properties of most of the PDS 456 companions detected with MUSE, we employ the Baldwin, Phillips, and Terlevich \citep[BPT][]{Baldwin81} diagram. As done above, we extract the same spectra from the datacube nuclear- and continuum-subtracted (Fig.~\ref{fig:BPTcompanions}, top panel). The emission lines are fit with Gaussian profiles, and the emission line ratios are derived and plotted on the BPT diagram (Fig.~\ref{fig:BPTcompanions}, bottom panel), with the color bar showing the distance in kpc from PDS 456. The BPT diagram indicates that M1, M4, and M5 fall within the AGN-dominated region, suggesting a significant contribution of AGN illumination. M2 is classified as HII regions, indicating ongoing star formation. Unfortunately, for the companion M3, information on the y-axis is unavailable, and we are unable to assign a specific classification.
For the sources M1, M4, and M5, we employ the formula from \cite{Keel12} to calculate the lower limit of the incident ionizing flux necessary for the observed \hb\ line emission, taking into account the distance of these companions from PDS 456. These estimates suggest that the ionizing flux originating from PDS 456 can explain the high ionization observed in the emission lines of all these companions. Therefore, their position on the BPT diagram is not attributed to the presence of an AGN.

\subsection{Luminosity and Mass of the diffuse ionized gas}\label{sec:gaspropwfm}

\begin{figure}[t]
   \begin{center}
   \includegraphics[height=0.52\textheight,angle=0]{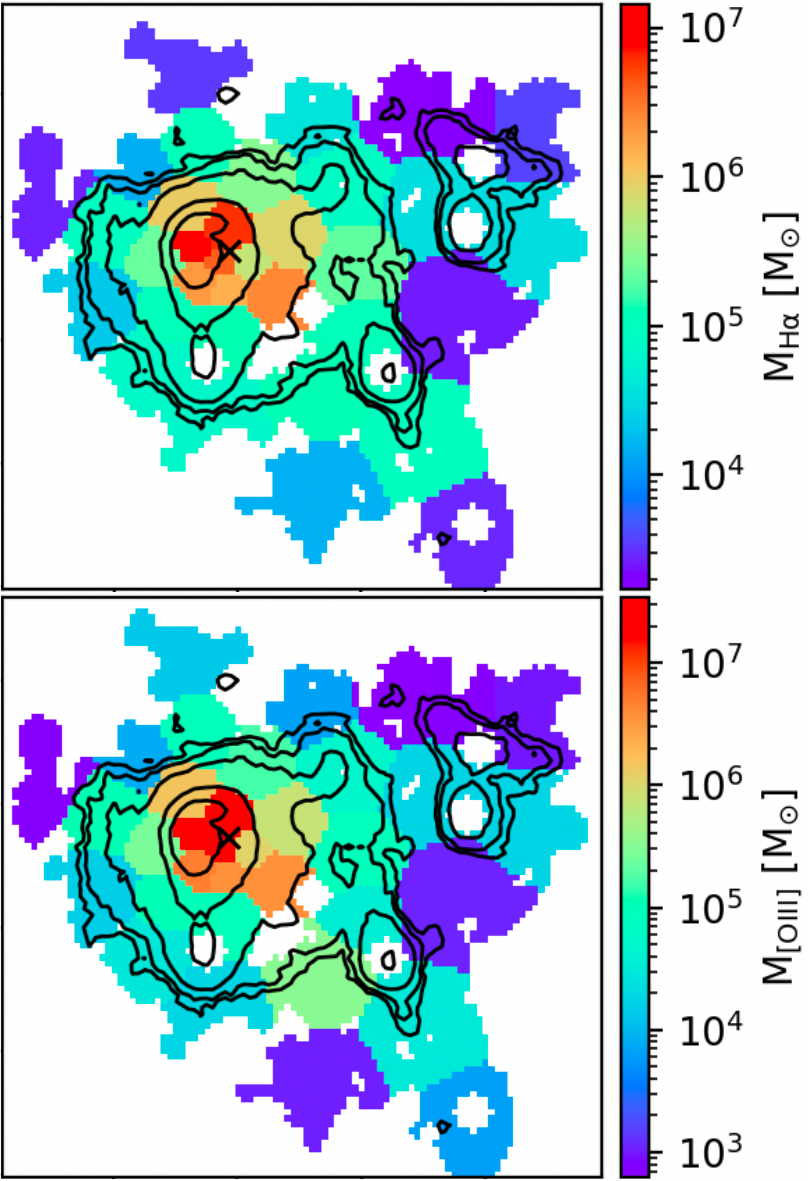}
   \caption{Maps of the extinction-corrected mass estimated from the \ha\ (top) and \oiii\ (bottom) emission, divided into Voronoi regions derived to reach a SNR$>3$ of the \hb\ line emission. The black thick contours in each map represent the SNR levels at 3, 5, 10, 30, and 50 of the \oiii\ line emission. The black cross marks the position of PDS 456.}
   \label{fig:Voronoi}
   \end{center}
\end{figure}

In this section, our primary objective is to conduct a comprehensive examination of the properties of the large-scale ionized gas surrounding PDS 456. To achieve this, we exclude the influence of galaxy companions by utilizing circular apertures with a 5-pixel diameter (i.e., approximately the PSF) centered on their positions. 

To estimate the gas properties, we utilize all available optical emission lines within Voronoi regions \citep[see][]{Venturi18,Molina22}. These regions are selected to have a SNR of \hb\ larger than 3, and we use the python tool \texttt{vorbin} for this purpose \citep{Cappellari03}. The \hb\ is crucial to estimate the extinction of the emission based on the \ha/\hb\ line ratio.

The maps of the ionized gas mass estimated from the extended \ha\ and \oiii\ emission in Voronoi regions (i.e. $M_{\ha}$ and $M_{\oiii}$, respectively) can be seen in Fig.~\ref{fig:Voronoi}. Within each Voronoi region, we extract the spectrum and perform a fit.
The fitting procedure involves modeling all emission lines using individual Gaussian profiles with identical widths, except for the \oiii\ doublet transition. In a few cases, we may observe a broader width in \oiii\ compared to the other emission lines. This discrepancy is attributed to the fact that \oiii\ is the most reliable tracer for ionized winds and shows a higher SNR, along with \ha, which is more affected by stellar absorption features. 
The centroid position of each Gaussian is set to have the same redshift, with an adjustment margin equal to the MUSE spectral resolution. The widths obtained from the best-fit models are deconvolved for the MUSE spectral resolution. Adopting these constraints in the modeling helps to minimize the impact of noise, instrumental and sky features, as well as any absorption that might affect the shape of individual emission lines \cite[see also][]{Veilleux23}.\\
 
We measure the electron density ($n_e$), color excess ($\rm E(B-V)$), ionized gas luminosity ($L_{\rm{ion}}$) and mass ($M_{\rm{ion}}$) derived from the \oiii\ and \ha\ emission lines.
To estimate $n_e$, we utilize the doublet \sii\ emission line ratio, as described in \cite{Osterbrock06}. The errors in $n_e$ are determined by applying a bootstrap algorithm for a sample of $n_e$ values obtained by randomly varying the intensities of the \sii\ emission lines within 1$\sigma$ error. In cases in which the SNR of the \sii\ doublet lines is less than 2.5, which tends to occur at larger distances from the quasar, we adopt a fixed value of $n_e = 150~\rm cm^{-3}$, which represents the minimum value calculated through the \sii\ line ratio in the outer regions. In this way, we find that $n_e$ is $\approx 600~\rm cm^{-3}$ at the center, decreasing to $\approx 150~\rm cm^{-3}$ in the outer regions. 
To correct for dust extinction in the observed flux, we estimate $\rm E(B-V)$ using the Balmer decrement \ha/\hb\ flux ratio. We apply the \cite{Calzetti00} attenuation law, assuming an intrinsic Balmer ratio \ha/\hb\ of 2.86 for gas at an electron temperature of $T_{\rm{e}} = 10^4 \rm ~K$ \citep{Osterbrock06}, and utilizing $\rm R_V = 3.12$. Near PDS 456, we obtain $\rm E(B-V)$ ranging from 0.2 to 1.1$~\rm{mag}$, decreasing at larger distances.
Then, we estimate the total luminosity, corrected for dust extinction, in \ha\ and \oiii\ to be $log(L_{\ha}/\rm{erg~s^{-1}}) = 42.47_{-0.31}^{+0.25}$ and $log(L_{\oiii}/\rm{erg~s^{-1}}) = 43.43 _{-0.30} ^{+0.26}$.

To estimate $M_{\rm{ion}}$, we use both the \ha\ and \oiii\ transition luminosity, corrected for dust extinction, according to the following formulas::
\begin{equation} \label{massaoiii}
	M_{\oiii} = 8 \times 10^7 \rm{M_{\odot}} \biggl( \frac{C}{10^{[O/H]-[O/H]_{\odot}}} \Biggr) \biggl( \frac{L_{[OIII]}}{10^{44}~\rm{erg~s^{-1}}} \Biggr) \biggl( \frac{n_e}{500~\rm{cm^{-3}}} \Biggr)^{-1}
\end{equation}
by \cite{Osterbrock06} \citep[or in][]{Carniani15,Bischetti17}, where $C$ is the clumpiness of the gas (i.e., $\rm \langle n_e ^2 \rangle / \langle n_e \rangle ^2 $), $\rm [O/H]$ is the oxygen abundance relative to hydrogen, and
\begin{equation} \label{massaha}
	M_{\ha} = 3.3 \times 10^8 \rm{M_{\odot}} \biggl( \frac{L_{H \alpha}}{10^{43}~\rm{erg~s^{-1}}} \Biggr) \biggl( \frac{n_e}{100~\rm{cm^{-3}}} \Biggr)^{-1} 
\end{equation}
by \cite{Nesvadba17} and \cite{Leung19}. 
In both cases, we assume a $T_e = 10^4~\rm K$ and solar metallicities. 
We correct $M_{\oiii}$ by multiplying it by a factor of three, as determined by \citetalias{Fiore17} due to the discrepancies between mass estimates obtained using equations \ref{massaoiii} and \ref{massaha}.
The total mass, as derived through the \ha\ and \oiii\ emission lines, is $2.75_{-2.26}^{+2.96}  \times  10^7~ \rm M_{\odot}$ and $8.69_{-6.74}^{+9.41} \times  10^7~ \rm M_{\odot}$, respectively.
These are two-three orders of magnitude lower than the virial dynamical mass of $\sim 10^{10}~\rm M_{\odot}$ derived within 1.3 kpc by \citetalias{Bischetti19}.

\begin{figure}[t]
   \begin{center}
   \includegraphics[height=0.3\textheight,angle=0]{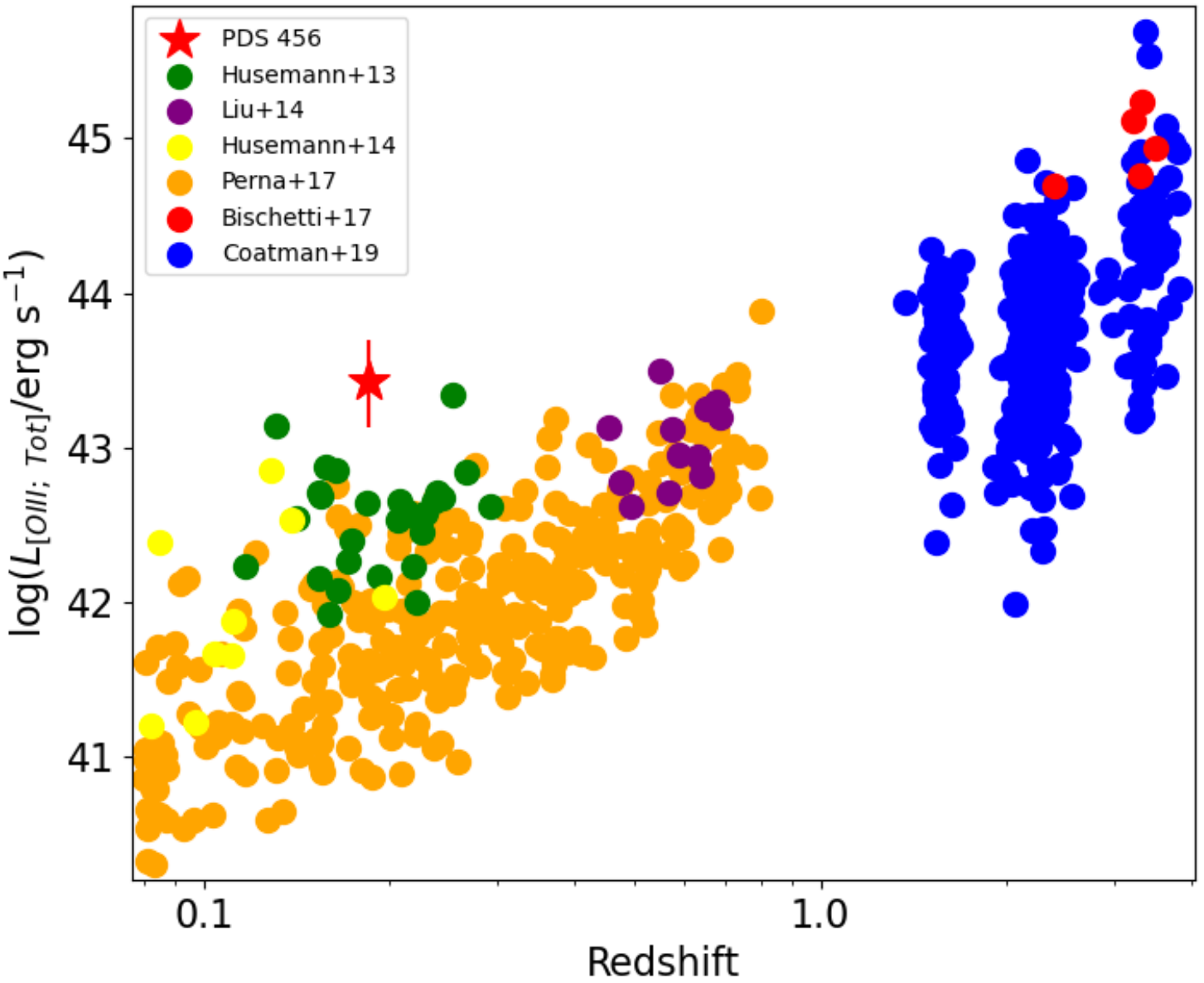} 
   \caption{Total observed \oiii\ luminosity ($L_{\oiii}$) of PDS 456 (red star) compared with those measured for other samples of Type 1 quasars in the redshift range $z\sim$0.1 to $\sim$4. The quasars at $z<$1 are from \cite{Husemann13} (green dots), \cite{Husemann14} (yellow dots), \cite{Liu14} (purple dots) and \cite{Perna17} (orange dots). Quasars at $z>$1 are from \cite{Coatman19} (blue dots) and \cite{Bischetti17} (red dots).}
   \label{fig:zL}
   \end{center}
\end{figure}

In Fig.~\ref{fig:zL} the total $L_{\oiii}$ measured for PDS 456 is compared to those derived for other samples of Type 1 quasars over a large redshift range ($z \sim $0.1-4). PDS 456 clearly lies at the brightest end of the $L_{\oiii}$ distribution for sources at $z \lesssim 1$. Moreover, its $L_{\oiii}$ is also consistent with that observed for the bulk of the quasar population at Cosmic Noon \citep[e.g.,][]{Coatman19} supporting the role of PDS 456 as a local counterpart of high-z quasars (and it is smaller by a factor of tens only when compared to that of hyper-luminous quasars like WISSH ones).


\subsection{The large-scale \oiii\ ionized outflow} \label{sec:pertquie}

\begin{figure*}[t]
   \begin{center}
   \includegraphics[height=0.44\textheight,angle=0]{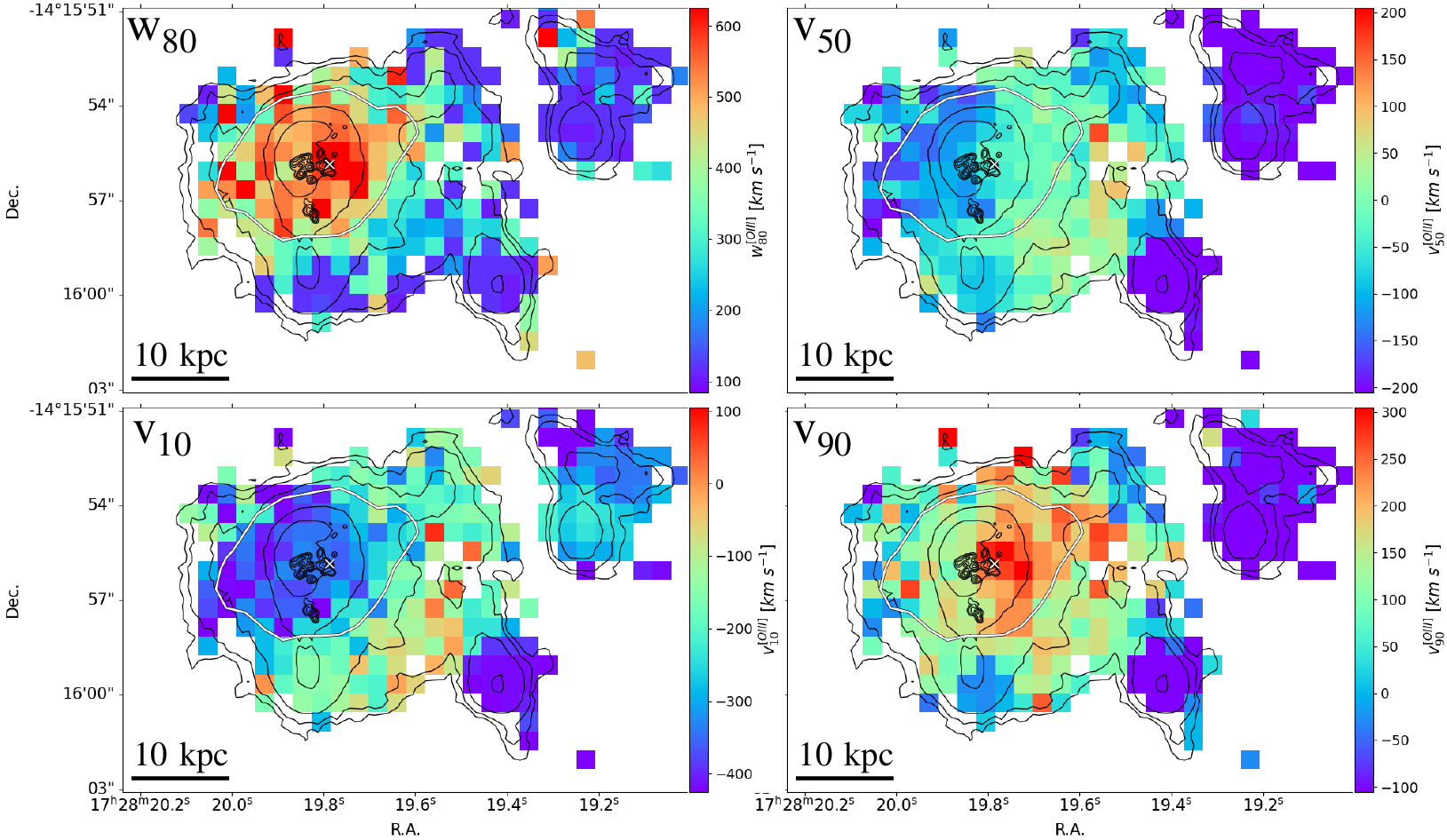} 
   \caption{Maps of $w_{80}$ (top left), $v_{50}$ (top right), $v_{10}$ (bottom left), and $v_{90}$ (bottom right), of the \oiii\ emission line. These maps provide an overview of the widths and velocity shifts from the entire \oiii\ profile. In each panel, the black thin contours represent the SNR levels shown in Fig.~\ref{fig:Voronoi}. The black thick contours in the center of the figure depict the molecular outflow, detected in \citetalias{Bischetti19}. These contours correspond to the emission levels of 1, 4, 8, and 10 mJy. The white cross symbol marks the position of the quasar. The white lines in each map represent 2$\rm \sigma$ Gaussian-smoothed contours that enclose the regions where $w_{80}$ exceeds 400$\rm ~km~s^{-1}$.}
   \label{fig:OutflowEmissionKin}
   \end{center}
\end{figure*}

In this section, we aim at revealing the presence of an ionized outflow based on the kinematics of the gas. 
To achieve an adequate SNR in the spectra, we bin the final nuclear- and continuum-subtracted datacube into $3 \times 3$ pixel regions\footnote{The analysis employing the Voronoi regions discussed in Sect.~\ref{sec:gaspropwfm} is not applicable in this context due to contamination in defining a region with specific kinematics. Instead, the Voronoi analysis is valuable for exploring the global properties of the gas, necessitating information from multiple lines.}, equivalent to $1.87~\times 1.87~ \rm kpc^2$.
Each spectrum extracted from this binned datacube is analyzed to model only the \oiii\ line emission, which is used as best proxy for the outflow. We adopt a single or double Gaussian model according with the Bayesian Information Criterion\footnote{The BIC statistic is calculated as the difference between $\chi^2 + k~ln(N)$, where N represents the number of data points and k is the number of free parameters in the model, for the single- and double-Gaussian models. Following this approach, a single-Gaussian profile is considered the best-fit model when the difference ($BIC_{\rm single} - BIC_{\rm double}$) is less than 10 as done in \cite{Vietri20}.} \citep[BIC;][]{Schwarz78}. 
From the best-fit models we derive the velocities maps: $v_{50}$, $v_{90}$, and $v_{10}$, representing the velocities containing a specific percentage of the total integrated flux of the emission line. Additionally, we calculate $w_{80}$, defined as $v_{90} - v_{10}$ \citep[see][]{Harrison14}. 
Commonly, the latter is used to trace kinematic properties associated with outflows \citep{Vega01,Collet16,Harrison16}. 

In Fig.~\ref{fig:OutflowEmissionKin}, the $w_{80}$ map reveals velocities exceeding $500 \rm ~km~s^{-1}$ in the vicinity of the PDS 456 core. These values gradually decrease to $\approx 300 \rm ~km~s^{-1}$ at larger radii and decrease further near the location of the companions. 
The $v_{50}$ map reveals negative velocities with $v_{50} < -100 \rm ~km~s^{-1}$ to the east of the quasar's position (Fig.~\ref{fig:OutflowEmissionKin}). On the west side, velocities are in the range $|v_{50}| < 50 \rm ~km~s^{-1}$, consistently with the maximum velocities observed in the \cott\ disk probed in ALMA data (\citetalias{Bischetti19}). 
Velocities for both the blue and red wings are determined using $v_{10}$ and $v_{90}$, respectively. These measurements indicate an east-west velocity gradient, with the blue wings extending to $-450 \rm ~km~s^{-1}$ in the east and the red wings reaching $250 \rm ~km~s^{-1}$ in the west. 

The kinematic maps highlight the following features: a notable velocity gradient in the east-west direction with negative to positive $v_{50}$ values, and perpendicular to the gradient of the compact molecular disk; a central $w_{80}$ exceeding $\rm \approx 500~km~s^{-1}$, more than twice the central velocity dispersion of the molecular gas (i.e. \sv $ = w_{80}/1.09/2.355$ for a Gaussian profile); and $v_{10} < -300~ \rm km~s^{-1}$, aligning with the extended \cott\ outflow represented by thick black contours in Fig.~\ref{fig:OutflowEmissionKin}, which is also blueshifted. 
These findings collectively suggest the presence of outflowing gas, which is further supported by the analysis of NFM data presented in Sect.~\ref{sec:NFMoution}. The latter reveal an unmistakable blueshifted outflow detected in \ha\ within $\approx$3 kpc from the center, aligned with the direction of the \cott\ outflow. 

This information can be used for pointing out the region where the outflow dominates in WFM MUSE data, essential for establishing the maximum projected distance of the \oiii\ outflow. As a rough estimate, we calculate an average \sv\ of the \ha\ outflow in NFM MUSE data of $\rm \sim 150 ~km~s^{-1}$ (see Fig.\ref{fig:SpectrumOutflows}). This corresponds to a $w_{80} = 400~ \rm km~s^{-1}$ and can be used as a reliable tracer for the \oiii\ outflow. Applying this criteria, we observe that the region where the ionized outflow dominates has a maximum projected size of $\approx 20~\rm kpc$, as shown by the white contours in each map in Fig.~\ref{fig:OutflowEmissionKin}.

In conclusion, our findings reveal compelling evidence of an outflow with a potential maximum projected distance of 20~kpc. This ionized outflow is characterized by $w_{80} \rm  \geq 500~km~s^{-1}$ in the center, and also exhibits blueshifted and redshifted velocities up to $\rm -450~km~s^{-1}$ and $\rm 250~km~s^{-1}$ to the east and west of the quasar, respectively.

\subsection{No Rotational pattern in the ionized disk}

ALMA data on PDS 456 reveal a compact (effective radius of 1.3 kpc) and nearly face-on (i$\sim$25 deg) molecular disk in \cott\ with a peak-to-peak velocity of $\rm 100 ~km ~s^{-1}$ along the NS direction (see Fig.~4 in \citetalias{Bischetti19}).
Under the assumption that the molecular and ionized phases of the gas disk share the same velocity gradient \citep{Levy18}, the kinematics of the ionized gas, probed with both \oiii\ and \ha, do not reveal an ionized counterpart to this disk. 
Instead, there is a gradient in velocities along the E-W direction (see the $v_{50}$ map in Fig.~\ref{fig:OutflowEmissionKin}), perpendicular to the velocity gradient of the molecular disk and indicative of the outflow kinematics.
Therefore, we suggest that the ionized disk is not detected, likely due to the rotational velocity of the disk being comparable to the MUSE spectral resolution, and the kinematics being dominated by the outflow.


\section{NFM observations}\label{sec:NFMobs}

The AO-NFM MUSE data provide us an unprecedented zoom-in of the central $\sim 24 \times 24~\rm kpc^2$ region of PDS 456 with a high spatial resolution ($ \sim 280~\rm pc$), reaching a surface brightness limit of $\approx 3 \times 10^{-19}~\rm{erg~s^{-1}~cm^{-2}~arcsec^{-2}}$ ($3 \sigma$).
This allows us to study in detail the morphology and kinematics of the ionized gas on the same scales of the molecular outflow \citepalias{Bischetti19}.

\begin{figure}[t]
   \begin{center}
   \includegraphics[height=0.37\textheight,angle=0]{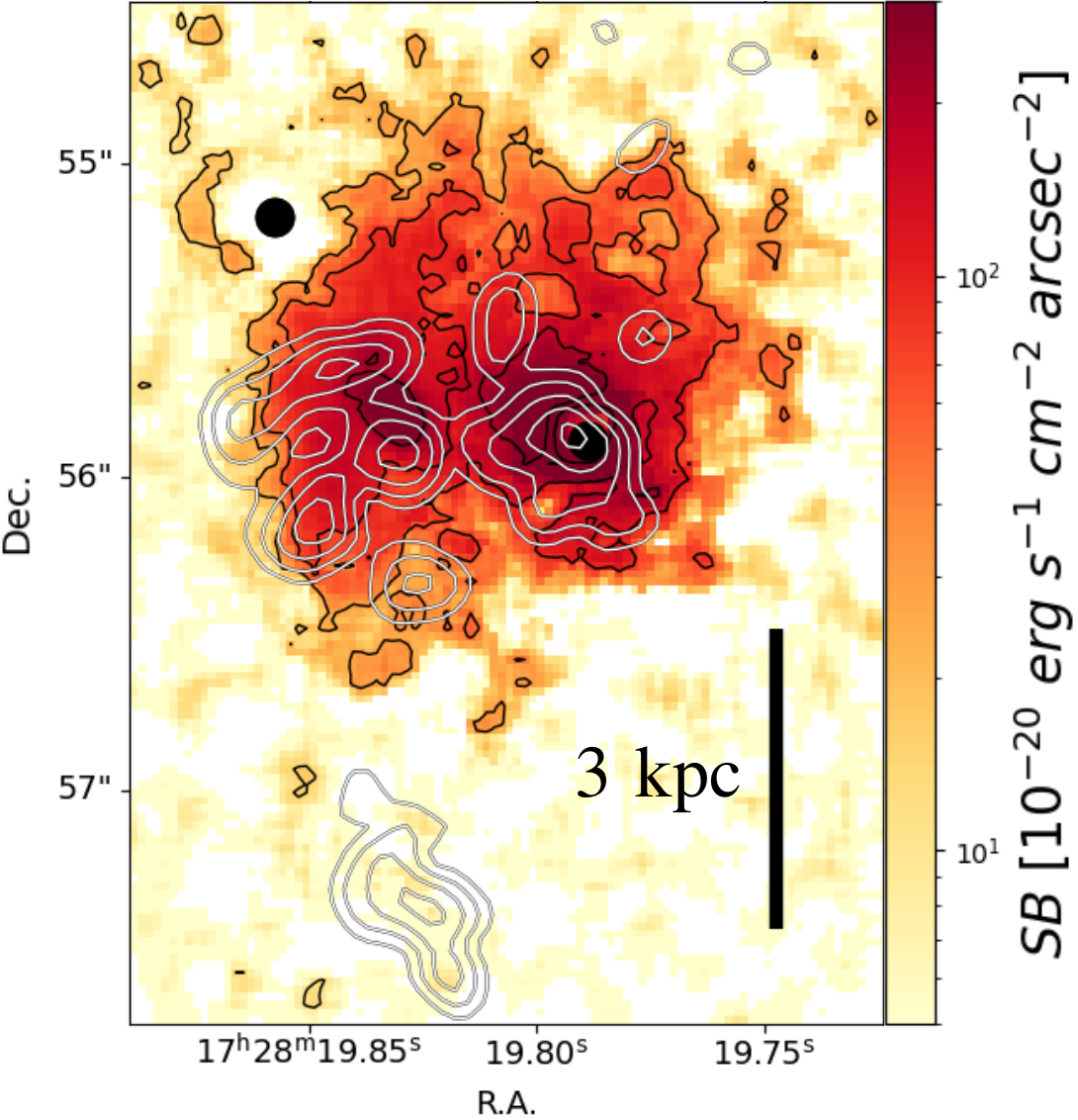}
   \caption{MUSE NFM Surface Brightness map of the \ha\ emission. The black contours show the SNR levels of the \ha\ emission of 3, 5, and 10. Grey contours show the morphology of the blueshifted \cott~molecular outflow (\citetalias{Bischetti19}), at the emission levels of 1, 4, 8 and 10 mJy. The black circles mark regions affected by significant residuals in the PSF subtraction due to the quasar and a nearby continuum source.}
   \label{fig:SBHaWeNFM}
   \end{center}
\end{figure}

\subsection{PSF and Continuum Subtraction}\label{sec:PSFsubNFM}

We use \texttt{CubExtractor} tools to subtract the PSF contribution of the quasar (including emission non-resolved from host-galaxy or NLR), and the continuum of the sources in the datacube. 
In particular, the \texttt{CubePSFSub} tool uses an empirical approach for the PSF subtraction at each wavelength based on pseudo-broad band images produced from the cube itself as described in detail in \cite{Cantalupo19}. This tool is specifically designed to extract faint and extended line emission on large scales from the bright continuum PSF of an isolated quasar assuming that it is the dominant source of continuum emission in the PSF area as it is in our NFM observations \citep[e.g.][]{Borisova16,Battaia19,Farina19,Cantalupo19,Travascio20}. 

As anticipated in Sect.~\ref{sec:reductionmuse}, the \oiii\ and \hb\ transitions are not covered as they fall in the \naid\ notch filter from the laser guide star, that contaminate that spectral region.
Hence, we perform the PSF and continuum subtraction on a datacube in which we mask the spectral regions including \ha, \nii\ doublet and \sii\ doublet emission lines.
The analysis to extract with \texttt{CubExtractor} the extended emission in AO-NFM MUSE data is the same used for WFM MUSE data described in Sect.~\ref{sec:ExtractionEmission}.
We extract the extended emission (in the \ha, \nii, and \sii\ transitions) by setting, in the \texttt{CubEx} function, a SNR threshold of 3 and a minimum number of connected voxels of 3000.

\subsection{High-Resolution view of the Ionized versus Molecular Spatially-Resolved Outflows} \label{sec:NFMoution}

The morphology of the \ha\ emission shown in Fig.~\ref{fig:SBHaWeNFM} (obtained from the AO-NFM observations) extends in a north-east direction with respect to the quasar, along the same direction of both the large-scale ionized and extended molecular outflows. The emission structure shows two peaks emission connected by a diffuse extended component. One peak is located close to the nucleus, while the other is situated 2~kpc east of the former, exhibiting a thick shell-like geometry. The morphology of this structure closely overlaps with that of the molecular outflow (\citetalias{Bischetti19}; grey contours), suggesting a multi-phase composition of the outflowing gas.
Due to the lower sensitivity of AO-NFM MUSE data, compared to that in the WFM MUSE data presented in the previous sections, we do not detect a similar elongated structure in the south direction. 

\begin{figure}[t]
   \begin{center}
   \includegraphics[height=0.55\textheight,angle=0]{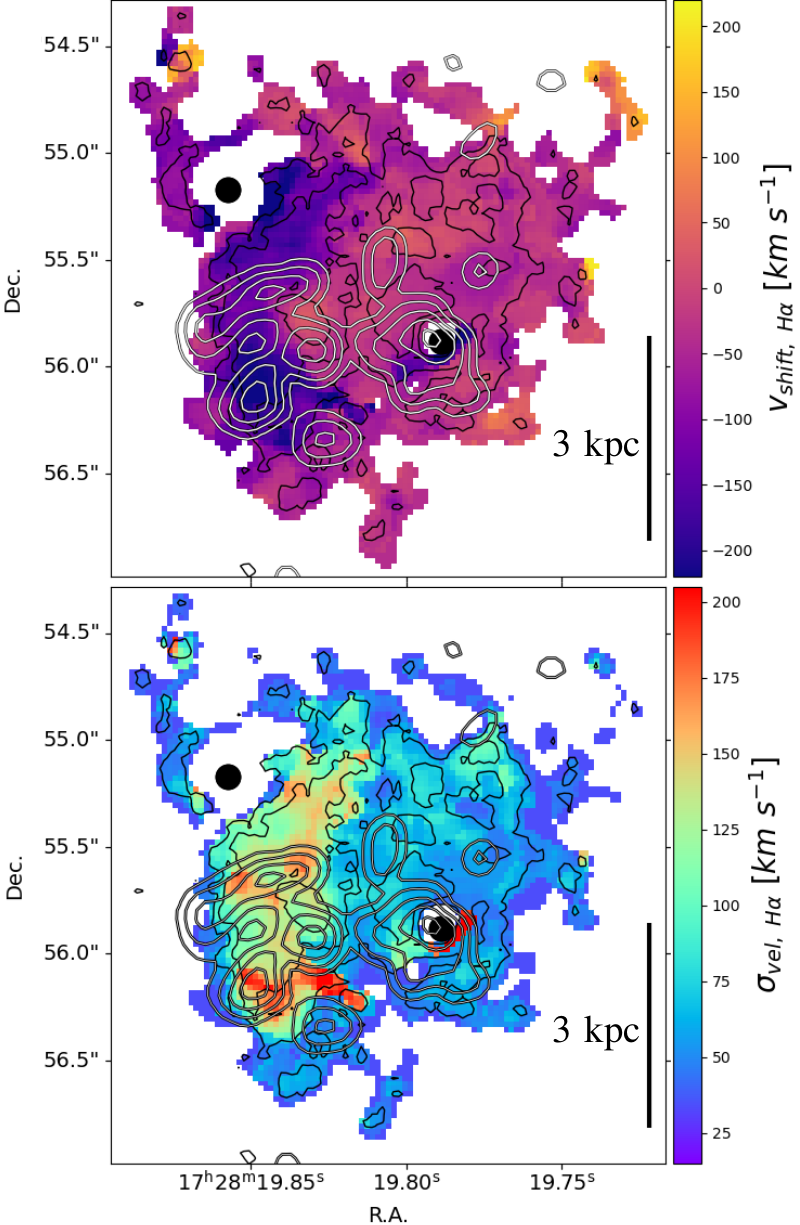}
   \caption{First (top) and Second (bottom) Moment of the flux distribution of the \ha\ emission obtained with the \texttt{Cube2Im} function in \texttt{CubExtractor}, representing the \vs\ and \sv\ maps. Gray contours show the morphology of the extended (up to $\approx 5 \rm~kpc$), blueshifted \cott~molecular outflow, detected by \citetalias{Bischetti19}, at the emission levels of 1, 4, 8 and 10 mJy. The black circles and contours are the same as in Fig.~\ref{fig:SBHaWeNFM}.}
   \label{fig:1}
   \end{center}
\end{figure}

Fig.\ref{fig:1} shows the maps of the 1st (top panel; \vs) and 2nd (bottom panel; \sv) moment of the \ha\ flux distribution obtained with the \texttt{Cube2Im} tool \citep[see][]{Borisova16}. The thick gray contours overlaid on the maps represent the morphology of the molecular outflow detected by \citetalias{Bischetti19}. 
In the $\approx 1.4 \rm~kpc$ thick shell-like region traced by the \ha\ emission, the \vs\ values are lower than $-150~\rm km ~s^{-1}$ and \sv\ values $\geq 120 \rm ~km ~s^{-1}$. These velocities differ from those present in the remaining regions, showing \vs\ between $-50$ and $50~ \rm km ~s^{-1}$ and \sv\ around $<100~\rm km~s^{-1}$. 
The spatial correlation between this shell and the molecular and large-scale ionized outflows suggests that this perturbed emission, up to \sv\ of $200 \rm~km ~s^{-1}$, is likely tracing an outflowing gas front.

\begin{figure}[t]
   \begin{center}
   \includegraphics[height=0.3\textheight,angle=0]{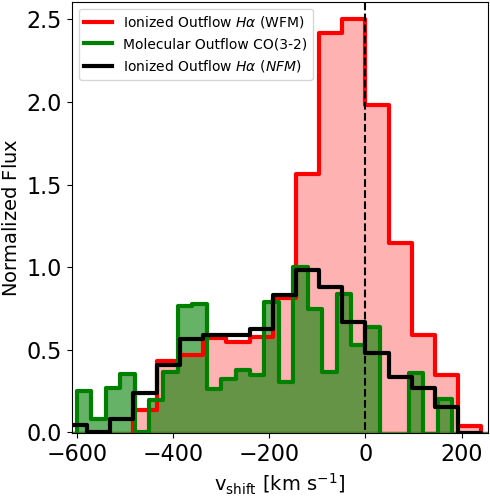}
   \caption{Spectra extracted from nuclear- and continuum-subtracted MUSE WFM (red; \ha) and NFM (black; \ha) datacubes, and from the ALMA data (green; \cott), in the region where the \ha\ outflow is observed in NFM MUSE data.}
   \label{fig:SpectrumOutflows}
   \end{center}
\end{figure}

Fig.\ref{fig:SpectrumOutflows} displays the spectrum of the \ha\ transition, zoomed in on the black \ha\ transition, extracted from the thick "shell"-like region with \sv\ higher than 90$\rm ~km~s^{-1}$ (maximum \sv\ in the center of the compact molecular disk). The spectrum of the \ha\ line emission (in red) from the WFM MUSE data, extracted from the same region, and the spectrum of the \cott\ outflow (in green) from \citetalias{Bischetti19}, are overlaid. The \ha-NFM and the \cott\ spectrum are normalized at the peak of the emission line, while the \ha-WFM spectrum is normalized to match its blueshifted wing with the \ha-NFM.
These spectra are plotted as a function of \vs\ with respect to the wavelength of the relative transition at the systemic redshift of the quasar. By comparing the maximum velocities reached by the wings of the emission lines, we observe identical \vs\ values for the \ha\ line emission in NFM and WFM MUSE data. We conclude that the molecular and ionized outflows share similar kinematics.
These similarities (i.e. extent, morphology, kinematics) strongly suggests the co-existence of different gas phases within these outflows.
Moreover, the morphology of the emission of the \cott\ molecular and the \ha\ ionized outflows may depend on the spatial variation of density, implying a clumpy structure of the gas \citep{Baron19}, although it could also be related to spatially varying emissivity or intrinsic dust absorption.

\begin{figure}[t]
   \begin{center}
   \includegraphics[height=0.48\textheight,angle=0]{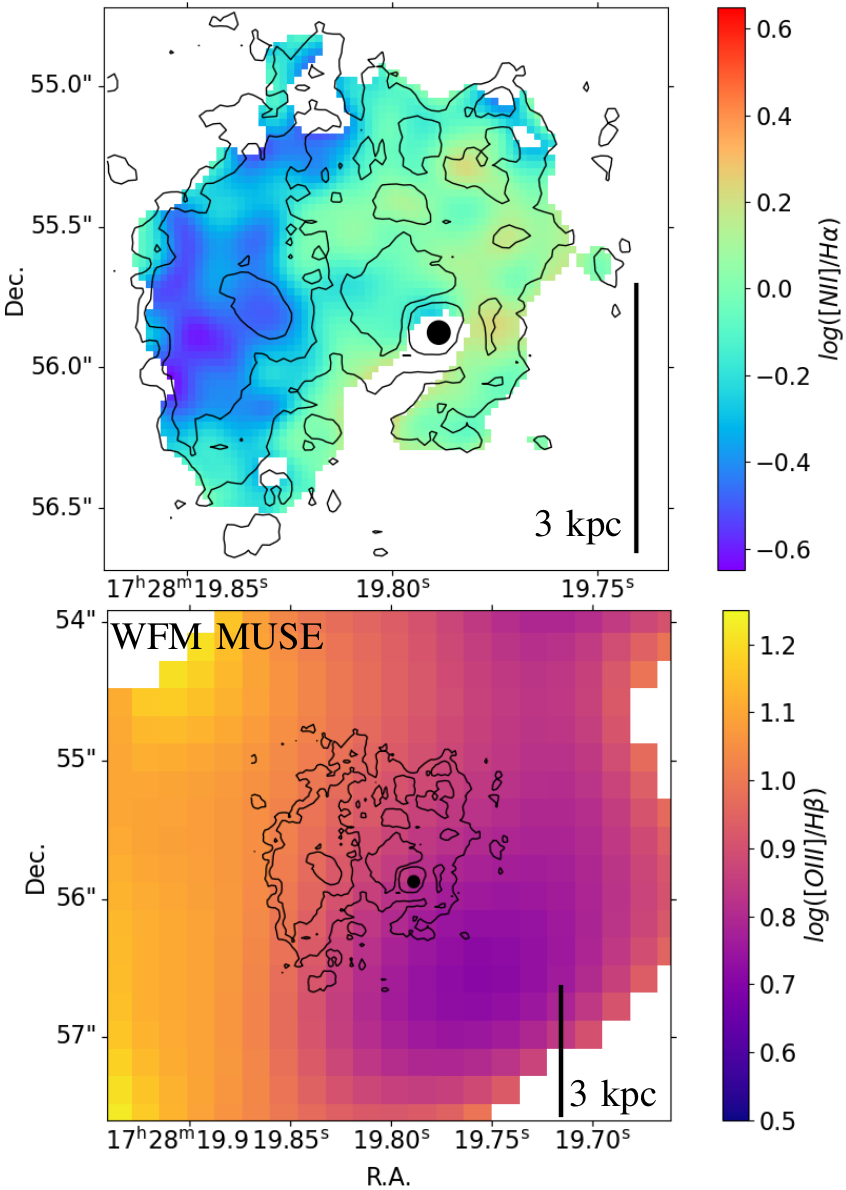}
   \caption{Top panel: \nii/\ha\ emission line ratio map, at $\rm SNR_{\ha} > 2.5$, whose pixel size is 5 times larger than the native pixel in NFM MUSE data, that corresponds to $\sim 1.3 \rm~FWHM_{PSF}$. The black dot marks the position of the central quasar, and has the FWHM size of the $\rm PSF_{NFM}$. Bottom panel: map of the \oiii/\hb\ line ratio estimated through the \oiii\ and \hb\ Optimally-Extracted NB images extracted from the WFM MUSE data with \texttt{CubExtractor}. The black contours represent the SNR levels of the \ha\ emission extracted from NFM MUSE data at 3, 5, and 10.}
   \label{fig:haniiratio}
   \end{center}
\end{figure}

\subsection{Ionization mechanisms of the outflowing gas}\label{sec:HaNFMemission}

In this section, we discuss the ionization mechanism of the outflowing gas in \ha, based on the spatially-resolved \nii/\ha\ and \oiii/\hb\ line ratios measured in NFM and WFM MUSE data, respectively, as key components in the diagnostic BPT diagram.

Fig.~\ref{fig:haniiratio} (top panel) shows a smoothed map of the \nii/\ha\ emission line ratio obtained from the respective optimally-extracted images derived with \texttt{Cube2Im}. Most of the emission associated with the outflow region exhibits a drop in the log(\nii/\ha)~($< -0.3$) with respect to the values in the entire map. Unfortunately, no other relevant emission lines fall in the spectral range covered by NFM MUSE data, while WFM MUSE data cannot provide line ratio estimates at the same spatial resolution of NFM data.
The low \nii/\ha\ line ratio weakens the hypothesis of shock excitation, which is typically associated with an increase in the \nii/\ha\ and \sii/\ha\ line ratios in the regions with highest values of \sv\ \citep[e.g.][]{Leung19}. This could be an indicator of star-formation occurring in the outflowing gas \citep[e.g.,][]{Maiolino17} or AGN ionization followed by the recombination of hydrogen atoms in a photon-dominated scenario, depending on the \oiii/\hb\ line ratio \citep[e.g.,][]{ODell09}.
To discern among these envisaged scenarios, the bottom panel in Fig.~\ref{fig:haniiratio} presents a map of the \oiii/\hb\ line ratio, derived from the optimally-extracted NB images obtained from WFM MUSE data with the \texttt{CubExtractor} tools by setting a SNR threshold of 3. We observe that the \oiii/\hb\ line ratio tends to increase in the region of the outflow detected in \ha\ emission in NFM MUSE data.

This, combined with the \nii/\ha\ line ratio, agrees with results of \cite{Hinkle19}, who suggest that an outflowing gas showing a larger \oiii /\hb\ line ratio and a smaller low-ionization line ratio (i.e. \nii / \ha, \sii / \ha, \oi / \ha) than those of the host galaxy ISM implies a difference in the ionization parameter. In addition, \cite{Mingozzi19} explain that outflowing gas in the NLR of Seyfert galaxies shows a low-\nii/\ha\ line ratio as it is directly illuminated by the AGN continuum. This may result from the \ha\ emission tracing matter-bound clouds within the outflow, which dominate the emission in the ionization cone \citep[see][for a discussion]{Mingozzi19}.

\begin{figure}[t]
   \begin{center}
   \includegraphics[height=0.5\textheight,angle=0]{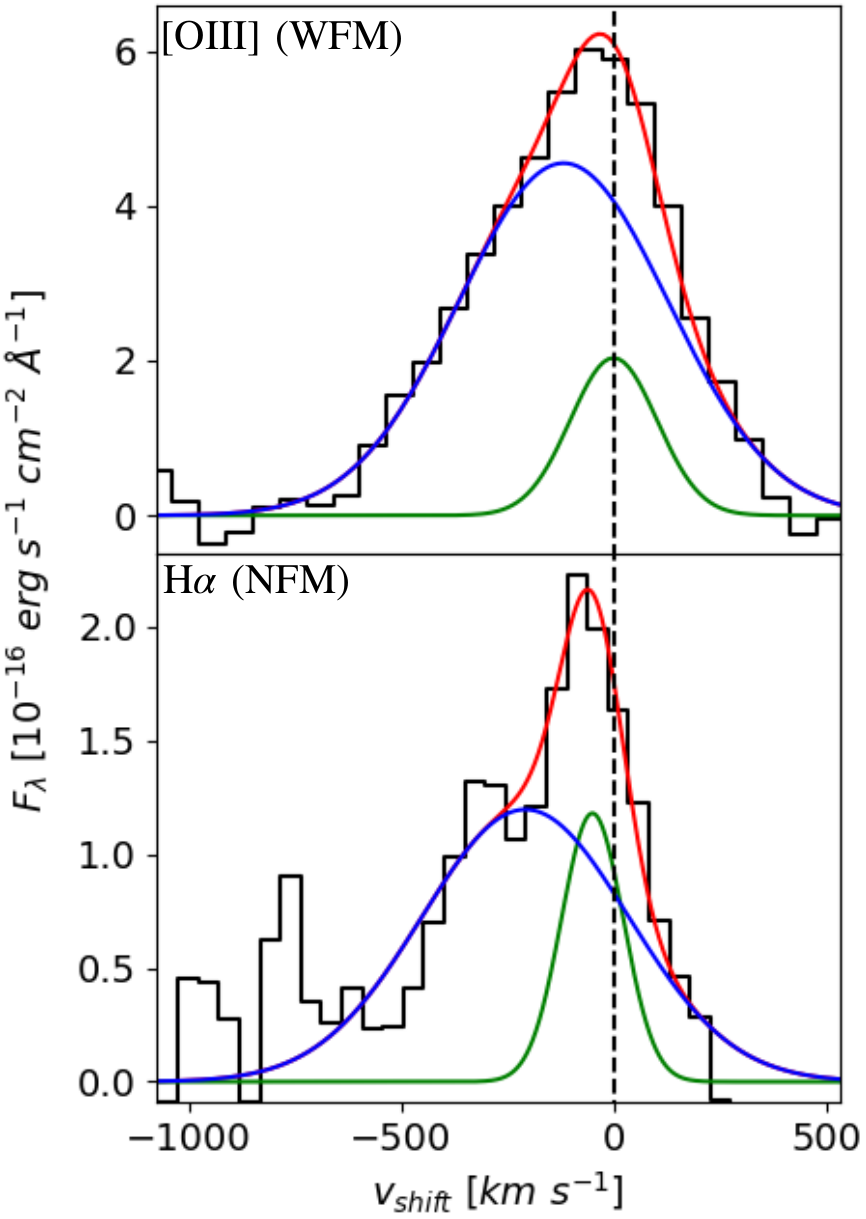}
   \caption{Spectra, in velocity units, extracted from the regions in which the outflow dominates the emission (see text for details), in WFM (top panel) and NFM (bottom panel) MUSE data. We report a zoom-in of the \oiii\ (top panel) and \ha\ (bottom panel) transitions. The red lines show the best-fit model consisting of the combination of two Gaussian components: a narrow one (green) and a broader one (blue), tracing the outflow.}
   \label{fig:fitOut}
   \end{center}
\end{figure}

\begin{table*}
\caption{List of the properties of the ionized (\ha, \oiii; top subtable) and molecular (\cott; bottom subtable in \citetalias{Bischetti19}) outflows in PDS 456. The numbers within the square brackets represent the minimum and maximum values within 1$\sigma$, respectively (see text for details). See text for details on the meaning of estimated properties.$^a$ For the ionized outflow, the reported values include $R_{\rm out}$ for the region traced by the \oiii\ and $\Delta R$ for the region traced by the \ha\ (see the text for details), while for the \cott\ outflow, the presented information includes the range of distances in which the molecular outflowing blobs are detected. $^b$ $v_{\rm out}$ for the the ionized and molecular outflows indicate the $v_{\rm max}$ (i.e. $v_{\rm{shift}} + 2  \sigma _{\rm vel}$) or the interval of velocities, respectively. The same ranges of radii and velocities for the molecular outflow are reported in Table 1 of \citetalias{Bischetti19}. They provide ranges for the radius and velocity of the outflow, as the properties of both the total and the extended outflow were derived by combining the characteristics of individual blobs. We consider only the molecular blobs spatially connected to the \ha\ outflow in the NFM MUSE observations when assessing the properties of the 1-3 kpc-scale molecular outflow.}
\label{tab:outen}
\bigskip
\centering\small\setlength\tabcolsep{1pt}
\begin{tabular}{c|c|c|c|c|c|c|c|c}
\hline
\hline
\multicolumn{9}{c}{Total Outflow (extended + unresolved)} \\
\hline 
  & $L_{\rm out}$ & $R_{\rm out}$ $ ^a$ & $v_{\rm out}$ $^b$  & $\langle n_e \rangle $ & $M_{\rm out}$ & $\dot{M}_{\rm out}$ & $\dot{E}_{\rm kin}$ & $\dot{P}_{\rm{out}}$ \\ [2pt]
  & $[10^{41}~\rm{erg~s^{-1}}]$ & $[\rm{kpc}]$ & $[\rm km~s^{-1}]$ & $\rm [cm^{-3}]$ & $[10^7~\rm M_{\odot}]$ & $[\rm M_{\odot}~yr^{-1}]$ & $[10^{42}~\rm{erg~s^{-1}}]$ & $[10^{34}~ \rm dyne]$ \\ [2pt]
\hline  
\oiii\ (WFM) & 38 [35-42] & 11.2 [6.5-11.4] & $593~[518-668]$ & $200$ &  2.3 [2.1-2.5]  &  3.8 [3.1-4.9]  &  0.44 [0.35-0.54] &  1.5 [1.2-1.7]  \\ [2pt]
 &   &   &   & 410 [334-497] &  1.1 [0.9-1.3] & 1.84 [1.47-2.37]  & 0.21 [0.17-0.26] & 0.71 [0.60-0.84]  \\ [2pt]
\hline 
 \cott\ &    & $\sim$1.8-5, $<$1.2  & $\rm [-1000,650]$  & & 25[16-28] & 290[180-760] & 40[7-58] & 120[78-320] \\ [2pt]
\hline 
\hline 
\multicolumn{9}{c}{1-3 kpc-scale Outflow (without unresolved component)}  \\ [2pt]
\hline
\ha\ (NFM) & 8~[6-9] & 1.4 [0.9-1.9] & 690 [540-840] & 500 [300-700]  & 0.47~[0.35-0.67] & 2.5~[1.9 - 3.1] & 0.38~[0.28-0.51] & 1.1~[0.9-1.3] \\ [2pt]
\hline 
\cott\ &  & 1.8-5 & $\rm [-1000,-250]$  & & 5.1~[4.5-5.7] & 42~[37-47] & 7.3~[1.3-10.6] & 19~[17-21] \\[2pt]
\hline
\hline
\end{tabular}
\end{table*}


\section{Energetics of the Ionized Outflow in PDS 456}\label{sec:Energy}

In this section, we estimate the properties ($M_{\rm out}$, \mdotout, \edotkin\ and \pdotout) of the ionized outflows in PDS 456 probed with \ha\ emission in NFM MUSE data (up to $\approx 3 \rm ~ kpc$ scales) and \oiii\ emission in the WFM MUSE data (up to $\approx 12 \rm ~ kpc$ scales).
First, we compare the properties of the extended ionized outflow probed with WFM MUSE data with those found in other AGNs. Most of them are reported in \citetalias{Fiore17} and updated in \citetalias{Bischetti19}, while others values can be found in \cite{Fluetsch19} and \cite{Speranza23}. \citetalias{Fiore17} assembled a sample of cold molecular outflows (CO, OH) and warm ionized outflows (\oiii, \ha, \hb) from AGN spanning a broad range of redshifts ($z<3$) and bolometric luminosities. Most of these AGN were studied using integral field unit (IFU) data.
Subsequently, given that the morphology and kinematics of the 1-3 kpc-scale (excluding the blob B, as detected by \citetalias{Bischetti19}, located 5 kpc south of PDS 456; see Fig.~\ref{fig:SBHaWeNFM}) blueshifted \cott\ and \ha\ outflows observed in the MUSE-NFM data suggest that these components are part of the same multi-phase outflow and are likely driven by the same past AGN feedback event (Sect.~\ref{sec:NFMoution}), we proceed to compare the properties and energy of these two outflow phases. This is crucial to explore the distribution of energy in different outflow phases and its correlation with distinct mechanisms that determine energy efficiency coupling.
We therefore calculate the integrated properties of the ionized outflow. To achieve this, we extract a spectrum from the WFM and NFM MUSE data in regions dominated by the outflow, defined as the region where $w_{80} \geq 400~\rm km~s^{-1}$ (see Sect.~\ref{sec:pertquie}), and the region where \sv\ is larger than $90 \rm ~km~s^{-1}$ (see Sect.~\ref{sec:NFMoution}), respectively. 
In both cases, we define the outflow based on a fitting model with two Gaussians: a narrow one (\sv $\approx 76 ~ \rm km~s^{-1}$) representing the systemic, non-outflowing gas, and a broader one (\sv $\approx 240 \rm ~km~s^{-1}$) tracing the outflow component.
From the best-fit model of the outflowing component, we derive some parameters for estimating additional outflow properties, and their associated 1$\sigma$ uncertainties are calculated using a bootstrap algorithm with $10^4$ iterations. This process involves propagating the errors obtained from the best-fit models of the initial spectra.

We derive the maximum velocity (i.e. $v_{\rm max} \equiv$  \vs\ $+ 2 \times$ \sv ; \citetalias{Fiore17}) and the observed integrated fluxes, that is corrected for dust extinction. The latter, characterized by $\rm E(B-V) \approx 0.52~\rm{mag}$, is estimated from the \ha/\hb\ line ratio observed in the central $5 \times 5 \rm ~kpc^2$ region in the WFM MUSE data, with the same method used in Sect.~\ref{sec:gaspropwfm}. This value is consistent with the results obtained by \cite{Reeves21} for PDS 456 and is also used for the extinction correction of the \ha\ outflow. 

The energetic properties that we derive for the outflow are dependent on $n_e$. Most of the studies of spatially-resolved kpc-scale outflows in AGN assume a uniform $n_e$ value, e.g. $< 500 \rm~cm^{-3}$ (\citealt{Kakkad16}, \citealt{Rupke17}, \citetalias{Fiore17}, \citealt{Carniani15}, \citealt{Bischetti17}) or $> 1000 \rm~cm^{-3}$ \citep{Muller11,Santoro18,Forster19,Baron19}. On the other hand, several studies show that $n_e$ of the outflows in AGNs can vary between $\sim 50 \rm~cm^{-3}$ and $10^6 \rm~ cm^{-3}$ \citep{Perna17,Kakkad18,Rose18,Baron19,Holden23,Venturi23}.
In addition, uncertainties on $n_e$ can be also large due to the limited data quality and the strong assumptions adopted in each method \citep{Baron19,Davies20,Revalski22}. 
\begin{figure}
   \begin{center}
   \includegraphics[height=0.4\textheight,angle=0]{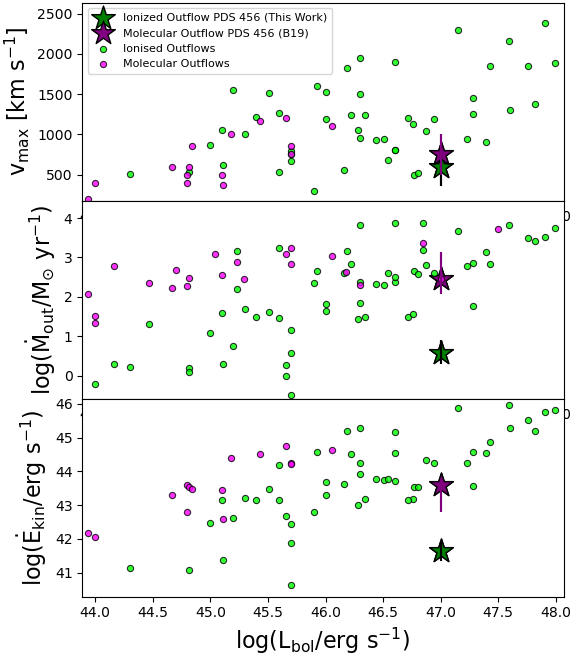}
   \caption{$v_{\rm max}$ (top panel), \mdotout\ (central panel) and \edotkin\ (bottom panel) of the molecular and ionized outflow in PDS 456 using star symbols, versus $L_{\rm bol}$. We compare these values with those of ionized and molecular outflows reported in \citetalias{Fiore17}, \citetalias{Bischetti19}, \citet{Fluetsch19} and \citet{Speranza23}. For the estimation of the ionized outflow's properties, we use an $n_e$ of $200~\rm cm^{-3}$ from \citetalias{Fiore17}.}
   \label{fig:LbolMoutEkin}
   \end{center}
\end{figure}
For a proper comparison of the \oiii\ outflow properties in PDS 456, as observed in WFM MUSE data, with those reported in \citetalias{Fiore17}, we use the same electron density value, $n_e = 200 \rm~cm^{-3}$, as employed in their study.
The mass of the \oiii\ outflow is derived using the equations~(\ref{massaoiii}) and, based on \citetalias{Fiore17}, the outflow mass estimated with the \oiii\ line is systematically smaller by a factor of 3 than that derived with \ha\ or \hb\ transitions, which are robust estimators of the mass. Therefore, we multiply the $M_{\oiii}$ in eq.~(\ref{massaoiii}) by a factor 3.
To calculate the mass outflow rate of the \oiii\ outflow we use the formula in \citetalias{Fiore17} assuming a cone geometry:
\begin{equation}
\dot{M}_{\rm out} = 3 \times \Biggl( \frac{M_{[OIII]} \times v_{\rm{max}}}{R_{\rm{out}}} \Biggr) 
\end{equation}
where $R_{\rm out}$ is the maximum radius up to which high velocity gas (i.e. $w_{80} > 400~ \rm km~s^{-1}$) is detected, following the approach used in the \citetalias{Fiore17} sample. Specifically, to take into account the uncertainties in the estimates of the radius of the bulk of the outflow, we compute the 50th percentile (the median value), as well as the 16th and 84th percentiles of the distances, weighted by the fluxes.
The main properties of the total outflow traced from the \oiii\ emission in WFM MUSE data are listed in Table~\ref{tab:outen} (top raw).

For the \oiii\ outflow detected in WFM MUSE data, we estimate $M_{\rm{out}} = (2.3 \pm 0.2) \times 10^7~\rm M_{\odot}$, $\dot{M}_{\rm{out}} = 3.6_{-0.8}^{+1.3}~ \rm M_{\odot}~yr^{-1}$ and $\dot{E}_{\rm kin} = (4.2 \pm 1) \times 10^{41}~\rm{erg~s^{-1}}$. 
In Fig.~\ref{fig:LbolMoutEkin} we report the plots $L_{\rm bol}$ versus $v_{\rm max}$ (top panel), \mdotout\ (central panel) and \edotkin\ (bottom panel).
The ionized outflow in PDS 456 exhibits a $v_{\rm max}$ that is consistent with the scatter of the relation in \citetalias{Fiore17}, at fixed $L_{\rm bol}$. The $v_{\rm max}$ of the total molecular outflow in PDS 456 \citepalias{Bischetti19} reaches values around $1000 \rm ~km~s^{-1}$, that is consistent with the values observed in the ionized outflow. Furthermore, it falls within the lower limit of the scatter observed in the population of molecular outflows. 
\mdotout\ and \edotkin\ of the ionized outflow in PDS 456 are $\approx$2-4 orders of magnitude lower than those estimated in the sample of \citetalias{Fiore17} at a similar $L_{\rm bol}$.
The properties of the molecular outflow in PDS 456 are slightly below, at the lower end, the extrapolation of the high-$L_{\rm{bol}}$ points for the molecular outflow in \citetalias{Fiore17}.
\mdotout\ and \edotkin\ of the total ionized outflow observed in PDS 456 are consistent with those of the well-studied outflow in the quasar XID2028 at $z \sim 1.59$ by using high-resolution JWST data \citep{Cresci23,Veilleux23}. The ionized outflow in XID2028 exhibits larger velocities ($\approx 1100~\rm  km~s^{-1}$) compared to that in PDS 456, but shows comparable extent ($\approx 10~\rm kpc$), $M_{\rm out}$ ($10^7-10^8~M_{\odot}$), and \mdotout\ ($3-6 ~\rm M_{\odot}~yr^{-1}$). However, the morphology of the ionized outflow and radio lobes in XID2028 suggests that they are connected, possibly indicating a jet-driven outflow.  On the contrary, the orientation and size of the ionized outflow and radio jet (up to 200 pc in size) in PDS 456 are uncorrelated, suggesting a radiative-driven outflow mechanism.
On the other hand, if we consider the \sii\ line ratio to estimate $n_e$ of the \oiii\ outflow \citep{Osterbrock06}, we obtain a value of approximately $400~\rm cm^{-3}$. Assuming this as the most accurate value, both \mdotout\ and \edotkin\ must be rescaled by a factor of two compared to those reported in Fig.~\ref{fig:LbolMoutEkin} (see Tab.~\ref{tab:outen}).

Thanks to the superb resolution of the NFM MUSE we are able to directly compare the ionized outflow traced by the \ha , and the \cott\ molecular outflows discovered by \citetalias{Bischetti19}, within $\approx$1-3 kpc from PDS 456 (see Fig.~\ref{fig:SBHaWeNFM}). For the \ha\ outflow, we derive $n_e = (500 \pm 200) \rm ~cm^{-3}$ using the \sii\ lines ratios \citep{Osterbrock06,Jiang19}, which is consistent with $n_e$ derived for the \oiii\ outflow observed in WFM MUSE data.  
The mass of this \ha\ outflow is estimated using the equation~(\ref{massaha}). To estimate the mass outflow rate, we adopt the formula from \citep{Husemann19}:
\begin{equation}
\dot{M}_{\rm out} = \Biggl( \frac{v_{\rm{max}}}{100~\rm{km~s^{-1}}} \Biggr) ~ \Biggl( \frac{M_{H \alpha}}{10^6~\rm{M_{\odot}}} \Biggr) ~\Biggl(\frac{100~\rm{pc}}{\Delta R} \Biggr) \rm ~\rm{M_{\odot}~yr^{-1}}
\end{equation}
which provides the mass rate within an outflowing shell of thickness $\Delta R$. 
The key properties of the \ha\ ionized outflow in the NFM MUSE data are summarized in Table~\ref{tab:outen} (bottom row).
$M_{\rm out}$, \mdotout\ and \pdotout\ for the \ha\ ionized outflow detected in NFM MUSE data, are about 1 order of magnitude lower than those for the molecular outflow.
The distinct gap between the properties carried by the ionized and molecular outflows is also evident for the total outflows as discussed in the previous section.
These discrepancies suggest that the ionized component of the outflow makes a negligible contribution in terms of energy. This finding is inconsistent with the expected energy equivalence based on the high-$L_{\rm bol}$ relation presented in \citetalias{Fiore17}.

\cite{Menci19} stressed that the mass outflow rate of outflows depends strongly on the direction of the outflow with respect to the plane of the disk. In particular, the mass rate of outflows is highest in the direction of the disk and decreases as the outflow moves away from the disk. This is due to the fact that the density of the gas in the disk is higher than that in the outflowing gas, and the mass outflow rate is proportional to the density of the gas. 
Based on this scenario, the observed difference in mass outflow rates between ionized and molecular outflows at the $\approx$1-3 kpc-scales, as well as the overall outflow, may be attributed to the different directions of these outflows in relation to the disk's plane. In this scenario, the molecular outflow could be triggered by nuclear winds impacting the ISM gas, while the ionized outflow might result from the radiative pressure of AGN photons on NLR clouds.
However, this hypothesis needs to be confirmed through further detailed analysis of the multi-phase outflow observed at the $\approx$1-3 kpc-scales.


\section{Scenarios for the expansion of multi-phase galactic-scale outflows} \label{sec:expansionOut}

We discuss here a variety of possible scenarios that could explain the properties of the multi-phase outflow in PDS 456.
\citetalias{Bischetti19} suggested that in a high-$L_{\rm bol}$ regime the ionized outflow may represent a significant fraction of the outflow mass, reducing the discrepancy between the low momentum boost measured from the molecular phase alone with the expectations for an energy conserving expansion of the nuclear UFO. Nonetheless, the momentum boost of the multiphase, galaxy-scale outflow of $(\dot{P}_{\rm mol} + \dot{P}_{\rm ion})/\dot{P}_{\rm rad} \approx 0.37$ is not consistent with an energy-conserving (i.e. non-radiative) expansion powered by the quasar, which predicts values in the range 5-10 \citep{Zubovas12}.

\subsection{Intermittent AGN phases scenario} \label{sec:7.2}

A possibility to reconcile the observations with an energy conserving expansion relies on the fact that the outflow might have been inflated during a previous AGN phase. Indeed, considering a velocity $\approx 300 \rm~km ~s^{-1}$, it takes the outflow $\sim$1~Myr to expand to the 0.3~kpc scale of the ``compact'' outflow, and $\sim$10~Myr to reach the 3~kpc size of the extended outflow. 
Even 1~Myr is longer than the expected typical duration of a single AGN episode, which is estimated by both statistical \citep{Schawinski15} and analytical arguments \citep{King15b} to be $<$0.1~Myr. So it appears likely that the outflows that are observed now were inflated by multiple previous AGN episodes. 
This scenario would also reconcile the kinetic power of the galactic-scale outflow in PDS (Fig.~\ref{fig:LbolMoutEkin}) with that expected from the \citetalias{Fiore17} relations, which reflect the overall quasar population and, thus, the long-term average of $L_{\rm bol}$.
The luminosity during a single AGN episode is not constant; rather, it probably decreases as the accretion disc is consumed, while feedback prevents more material from reaching the disc efficiently \citep{King07}. \cite{Zubovas20} showed that in such cases, outflow properties tend to correlate better with the long-term average AGN luminosity than the instantaneous one. Given that the current luminosity of PDS 456 is $L_{\rm bol} \simeq 0.7~L_{\rm Edd}$ \citep{Reeves00,Nardini15} it is very likely that the long-term average luminosity is much lower. The UFO revealed in X-rays, on the other hand, probably evolves on sub-year timescales and so closely tracks the present-day luminosity. Assuming that the nuclear wind always had kinetic power equal to $\sim 20 \% ~L_{\rm bol}$, as today, a long-term average luminosity 10-20 times lower than the present-day value would bring the outflow properties in line with the expectation of an energy-driven scenario. Such a low average luminosity is easy to achieve: it requires the product of AGN duty cycle over the past 1-10~Myr and the average Eddington factor during an episode to be of order 0.035-0.07. The AGN duty cycle is defined in \cite{Zubovas20} as $\rm \delta _{\rm{AGN}} = t_{\rm{ep}}/t_r$, where $\rm t_{\rm{ep}}$ represents the duration of the AGN episode, which is a measure of how long the AGN remains in an active state, while $\rm t_r$ refers to the recurrence time scale, which determines the frequency of AGN activity. By definition, the average luminosity during an episode is at least a factor few below Eddington (i.e. $L_{\rm{AGN}} > 0.01 \times L_{\rm{Edd}}$), so the duty cycle has to be somewhat higher than the population average of $\sim$0.07 to achieve this. Moreover, \cite{Zubovas22} suggested that AGN episodes may be clustered hierarchically in time, with longer phases lasting $10^7~\rm yr$ \citep[as suggested by][]{Hopkins05} during which multiple AGN events, each lasting $\sim 10^5~\rm yr$, occur.

\subsection{Radiation-pressure driven outflow scenario} \label{sec:7.3}

Yet another possibility, as suggested by \citetalias{Bischetti19}, is that radiation pressure on dust has triggered these outflows in PDS 456, implying a value of the momentum load of galactic-scale outflow around unity \citep{Ishibashi14,Costa18}. This scenario might be supported by the detection of outflowing emission traced by blueshifted high-ionization emission lines in the mid-IR (i.e. $15.56 ~\rm \mu m$ [\ion{Ne}{III}] and $14.32 ~\rm \mu m$ [\ion{Ne}{V}]) with a velocity larger than those found for \oiii\ and \ha\ and, therefore,  related to an inner, highly-extincted ionized wind component \citep[e.g.,][]{Spoon09}; or detection of dust emission co-spatial with the multi-phase outflow.

\subsection{Star-formation and radio-jet driven scenario} \label{sec:7.4}

Both an origin of the multi-phase outflow in PDS 456 due to star formation activity or one linked to the power of a radio jet seem to be unlikely. Specifically, \citetalias{Bischetti19} excluded a dominant starburst contribution to the outflow acceleration since the kinetic power of the molecular outflow is significantly larger than that expected for a wind triggered by a star formation activity with a $SFR \simeq 50 - 80~\rm M_{\odot}$ as observed in PDS 456.

\cite{Yang21} detected a complex radio structure comprising a radio core, a radio jet, and a diffuse component. These radio components extend to distances of $\sim 370~pc$ from the quasar, with luminosities of $log(L_{\nu} \nu/ \rm{erg~s^{-1}}) = 40$ ($\rm \nu = 1.66~ GHz$), and $log(L_{\nu} \nu/\rm{erg~s^{-1}}) = 39.3$ for the radio core and jet, respectively. A strong correlation between the galactic-scale outflows and the radio jets seems unlikely because the radio luminosity is more than two orders of magnitude lower than the kinetic power associated with the galactic-scale outflows, although based only on the luminosity at $\rm \nu = 1.66~ GHz$. Moreover, the scales and orientation of the radio emission (see Fig. 2 in \citet{Yang21}) are uncorrelated with those of the ionized and molecular outflows (Fig.~\ref{fig:OutflowEmissionKin} and \ref{fig:1}), suggesting that they are not related.


\section{Summary and Conclusions} \label{sec:summconc}

We carried out the analysis of the first VLT/MUSE WFM and AO-NFM observation of the nearby ($z=0.185$), luminous ($L_{\rm bol} \sim 10^{47} \rm erg~s^{-1}$) quasar PDS 456. These data provided us with the opportunity of mapping with unprecedented detail the ionized emission from a luminous quasar with a multi-scale, multi-phase outflow.

Specifically, with the WFM MUSE data, we were able to investigate the environment and a kpc-scale outflow. The AO-NFM MUSE data offered an unmatched spatial resolution of $\sim$280~pc, enabling us to accurately study in detail the morphology and kinematics of the ionized outflow and compare it with those of the molecular outflow, with equally resolution and at the same scales. 
Our analysis provided the following key results:

\begin{enumerate}

\item PDS 456 resides in a complex environment characterized by the diffuse emission of ionized gas extending up to a maximum projected distance of $\sim 46~ \rm kpc$, which traces the CGM and the presence of multiple line-emitting companion galaxies within $\sim$30-40 kpc (See Fig.~\ref{fig:WFMmaps}). We estimate the mass of the emitting ionized gas, through the \oiii\ and \ha\ emission lines, $\sim$3 and $\sim$9$\times 10^7~\rm M_{\odot}$, respectively (see Sect.~\ref{sec:gaspropwfm}). These findings match with theoretical and observational expectations suggesting that hyper-luminous quasars live in over-dense regions in terms of companion galaxies \citep{Wagg12,Decarli17,Trakhtenbrot17,Bischetti21,Nguyen20,Perna23} and reservoir of gas \citep{Bowen06,Prochaska09,Farina13}. 
 
\item We also discover the existence of an ionized outflow ($v_{max} \sim \rm 600~km~s^{-1}$), mainly detected in \oiii, and characterized by a maximum projected size of $\approx 20~\rm kpc$ and a velocity gradient along the east-west direction (see Sect.~\ref{sec:NFMoution}; Fig.~\ref{fig:OutflowEmissionKin}). 

\item \mdotout\ and \edotkin\ of the \oiii\ ionized outflow in PDS 456 are 2-4 orders of magnitude less than those derived in objects with similar $L_{\rm bol}$ in literature \citep[\citetalias{Fiore17,Bischetti19};][]{Fluetsch19,Speranza23}, while the measured value of $v_{\rm max} \approx 600 \rm ~km~s^{-1}$ is consistent with those typically found in other quasars (see Fig.~\ref{fig:LbolMoutEkin}). 

\item The total multiphase outflow expands in a regime not consistent with energy conservation, contrary to what might be expected given its scale \citep{King15}. In Sect.~\ref{sec:7.2}, we discuss in detail the intermitted AGN phase scenario, that could provide an explanation for this finding.

\item The spatially resolved \ha\ emission in the AO-NFM MUSE data partially traces the same \oiii\ outflow detected with WFM MUSE data down to $\approx 1-3~$kpc-scales from the quasar. The high resolution enables us to reveal a thick shell-like geometry (see Fig.~\ref{fig:SBHaWeNFM}).

\item The remarkable similarity in morphology, direction, and kinematics between the \ha\ outflow detected in NFM MUSE data and the extended \cott\ molecular outflow strongly suggests that these two components belong to the same multi-phase outflow and could be driven from the same past AGN feedback episode (see Figs.~\ref{fig:1} and \ref{fig:SpectrumOutflows}).
 
\item The molecular phase dominates the kinetic power of the multi-phase, galaxy-scale outflow, with a $\dot{M}_{\rm mol}$ and a \pdotout\ that is about 1 order of magnitude larger than that of the ionized outflow. 

\end{enumerate} 

In conclusion, our analysis of the MUSE data unveils novel and important details on the complex interplay between the different phases of AGN-driven outflows extending up to $\sim 20 \rm ~ kpc$ (i.e. interacting with the CGM), and the richness of circum-galactic environment of PDS 456. This may help in planning and performing future multi-band investigations of the properties of the luminous quasars shining at cosmic noon. 

We highlight that deeper MUSE observations in both WFM and NFM modes would allow us to enhance and broaden our analyses.
Specifically, these WFM observations could reveal a more extensive CGM structure associated with PDS 456 across all optical transitions. Data with an larger SNR may help in identifying a fainter wing of the \oiii, suggesting the presence of a less dominant or more obscured outflow at higher velocities.
In addition, deeper MUSE NFM observations would be crucial for investigating the presence of more extended quiescent and outflowing ionized gas.
This is very important for comparing the differences in clumpiness between the ionized and molecular phases of the gas and improving our ability to produce spatially resolved $n_e$ maps of the outflow, thanks to the high SNR maps of \sii\ emission lines.

Furthermore, JWST/NIRSpec-IFU data (filter G235H/F170LP) and MIRI-IFU data (filter F770W) have been recently acquired. These data are crucial to investigate both hot molecular and highly-extincted ionized gas, potentially revealing additional phases of the outflow in PDS 456 \citep{Spoon09,Bianchin22}. Additionally, the rest-frame near-IR $\rm H_2$ emission line can be used as a valuable tracer for shocks and contributing to a clearer understanding of the main expansion mechanism in the shell.

\begin{acknowledgements}
We thank the referee for the careful reading of the manuscript and the helpful comments.
Based on observations made with ESO Telescopes at the La Silla Paranal Observatory as part of the ESO programme ID 0103.B-0767(A-B).
This project was supported by the European Research Council (ERC) Consolidator Grant 864361 (CosmicWeb) and by HORIZON2020: AHEAD2020-Grant Agreement n. 871158.
Support from the Bando Ricerca Fondamentale INAF 2022 Large Grant "Toward an holistic view of the Titans: multi-band observations of $z > 6$ quasars powered by greedy supermassive black-holes" is acknwoledged. EP and GV acknowledges the PRIN MIUR project ‘Black hole winds and the baryon life cycle of galaxies: the stone guest at the galaxy evolution supper’, contract number 2017PH3WAT).
EP and FT acknowledge funding from the European Union - Next Generation EU, PRIN/MUR 2022 2022K9N5B4.
GC acknowledges the support of the INAF Large Grant 2022 “The metal circle: a new sharp view of the baryon cycle up to Cosmic Dawn with the latest generation IFU facilities”. 
MB acknowledges support from INAF under project 1.05.12.04.01 - 431 MINI-GRANTS di RSN1 "Mini-feedback"  and support from UniTs under project DF-microgrants23 "Hyper-Gal".
SC and GV acknowledge support by European Union’s HE ERC Starting Grant No. 101040227 - WINGS.
CRA acknowledges support from projects ``Feeding and feedback in active galaxies'', with reference PID2019-106027GB-C42, funded by MICINN-AEI/10.13039/501100011033, and ``Tracking active galactic nuclei feedback from parsec to kiloparsec scales'', with reference PID2022-141105NB-I00. MP acknowledges support from the research project PID2021-127718NB-I00 of the Spanish Ministry of Science and Innovation/State Agency of Research (MCIN-AEI/10.13039/501100011033).
\end{acknowledgements}

\newpage

\bibliography{andrea}{}
\bibliographystyle{aa}

\newpage

\begin{appendix}

\section{Best-fit \naid\ absorption transition}\label{app:abs}

\begin{figure}[t]
   \begin{center}
   \includegraphics[height=0.55\textheight,angle=0]{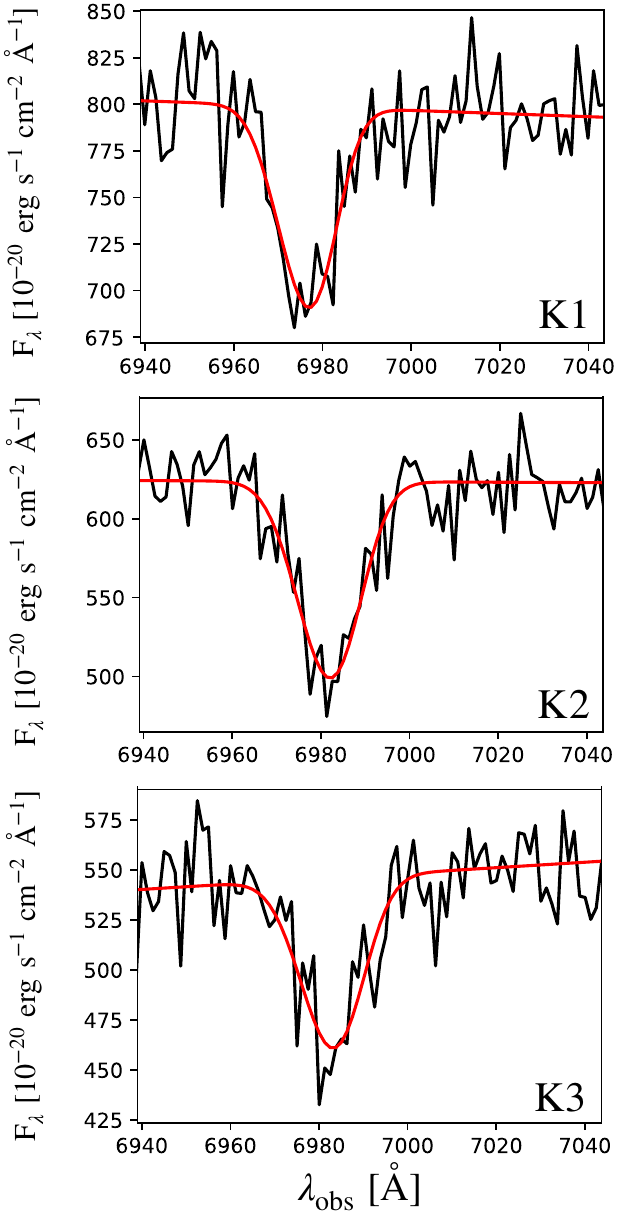}
   \caption{Spectra extracted from 2-pixel radius circles centered on the sources K1, K2, and K3, zoomed in on the \naid\ absorption line. The red lines is the best-fit models of the \naid\ absorption features.}
   \label{figapp:abs}
   \end{center}
\end{figure}

Fig.~\ref{figapp:abs} shows the best-fit models of the \naid\ absorption lines observed at the position of the K1, K2 and K3 sources in WFM MUSE data. As detailed in Sect.~\ref{sec:companions}, we extract the spectra at the sources position in 2-pixel radius circles and we apply the fitting model to follow:
\begin{equation}
	I(\lambda) = I_{\rm{em}} \times (1 - C_f \times [1 - e^{ - \tau _0^{-(\lambda - \lambda _K)^2/(\lambda _K b / c)} -2 \tau e^{-(\lambda - \lambda _H)^2/(\lambda _H b / c)}}])
\end{equation}
where $I_{\rm{em}}$ is the normalization parameter, $\lambda_K$ and $\lambda_H$ are the wavelengths of the sodium at 5891$~\AA$ and 5896$~\AA$ respectively, $C_f$ is the covering factor, $\tau_0$ is the optical depth at the central $\lambda$, $b$ is the Doppler parameter, and $c$ is the speed of light. 
This model is widely used in literature \citep[e.g.,][]{Rupke05,Sato09,Perna20} and provides a physically-motivated description of the absorption features, which can be used to derive column density and mass of the absorbing gas (Travascio et al. in prep.).
The redshift estimated in this way for the sources K1 and K3, are consistent with those derived through molecular transitions with ALMA by \citetalias{Bischetti19}.

\end{appendix}

\end{document}